\documentclass[a4paper,fleqn,usenatbib]{mnras}

\usepackage{newtxtext,newtxmath}
\usepackage[T1]{fontenc}
\usepackage{ae,aecompl}

\usepackage{graphicx}	% Including figure files
\usepackage{amsmath}	% Advanced maths commands
\usepackage{amssymb}	% Extra maths symbols

\newcommand{\Eexc}{$E_{\rm exc}$}
\newcommand{\Teff}{T_{\rm eff}}
\newcommand{\logg}{\rm log~ g}
\newcommand{\kms}{km\,s$^{-1}$}
\newcommand{\eps}[1]{\log\varepsilon_{\rm #1}}
\newcommand{\eu}[5]{\mbox{$#1\,^#2{\rm #3}^{#4}_{\rm #5}$}}
\def\ione{\,{\sc i}}
\def\ii{\,{\sc ii}}
\def\iii{\,{\sc iii}}
\def\iv{\,{\sc iv}}

\title[Non-LTE line formation for Si\ione-\ii-\iii\ in A-B stars]{Non-LTE line formation for Si\ione-\ii-\iii\ in A-B stars and the origin of Si\ii\ emission lines in $\iota$ Her}

\author[L. Mashonkina]{
Lyudmila Mashonkina,$^{1}$\thanks{E-mail: lima@inasan.ru}
\\
$^{1}$Institute of Astronomy of the Russian Academy of Sciences, Pyatnitskaya st. 48, 119017, Moscow, Russia
}

\date{Accepted XXX. Received YYY; in original form ZZZ}

\pubyear{2020}

\begin{document}
\label{firstpage}
\pagerange{\pageref{firstpage}--\pageref{lastpage}}
\maketitle

% Abstract of the paper
\begin{abstract}
A comprehensive model atom was developed for Si\ione -\ii -\iii\ using the most up-to-date atomic data available so far. Based on the non-local thermodynamic equilibrium (NLTE) line formation for Si\ione, Si\ii, and Si\iii\ and high-resolution observed spectra, we determined the NLTE abundances for a sample of nine unevolved A9 to B3-type stars with well determined atmospheric parameters. For each star, NLTE reduces substantially the line-to-line scatter for Si\ii\ compared with the LTE case and leads to consistent mean abundances from lines of different ionisation stages. In the hottest star of our sample, $\iota$~Her, Si\ii\ is subject to overionisation that drives emission in the lines arising from the high-excitation doublet levels. Our NLTE calculations reproduced 10 emission lines of Si\ii\ observed in $\iota$~Her. The same overionisation effect leads to greatly weakened Si\ii\ lines, which are observed in absorption in $\iota$~Her. Large positive NLTE abundance corrections (up to 0.98~dex for 5055~\AA) were useful for achieving consistent mean abundances from lines of the two ionisation stages, Si\ii\ and Si\iii. It was found that the NLTE effects are overestimated for the Si\ii\ 6347, 6371\,\AA\ doublet in $\iota$~Her, while the new model atom works well for the cooler stars. At this stage, we failed to understand this problem. We computed a grid of the NLTE abundance corrections for lines of Si\ione, Si\ii, and Si\iii\ in the model atmospheres with effective temperatures and surface gravities characteristic of unevolved A-B type stars.

%The abstract should briefly describe the aims, methods, and main results of the paper.
%It should be a single paragraph not more than 250 words (200 words for Letters).
%No references should appear in the abstract.
\end{abstract}


%%%%%%%%%%%%%%%%%%%%%%%%%%%%%%%%%%%%%%%%%%%%%%%%%%

%%%%%%%%%%%%%%%%% BODY OF PAPER %%%%%%%%%%%%%%%%%%

\section{Introduction}

Silicon is one of the most abundant metals. Analyses of the silicon lines in A- to mid B-type stars are important for addressing several problems. Such stars are young and can serve as indicators for present-day cosmic abundances. Using a sample of unevolved early B-type stars in nearby OB associations and the field and the non-local thermodynamic equilibrium (non-LTE = NLTE) line formation for Si\iii, \citet{2008ApJ...688L.103P} and \citet{2012A&A...539A.143N} establish the Cosmic Abundance Standard and prove that the mean stellar silicon abundance agrees within 0.01~dex with the solar one. \citet{2009AA...503..945F} investigate whether the chemical composition of the solar photosphere can be regarded as a reference also for early A and late B-type stars. For each of their three stars, the LTE analysis led to quite discordant abundances from lines of different ionisation stages of silicon. \citet{2009AA...503..945F} claim the real need of calculations based on the NLTE line formation for Si\ione\ - Si\ii\ - Si\iii\ in A to mid B-type stars.

Exactly in this range of spectral types, spectroscopic peculiarities (chemical abundance peculiarities, stratification, Zeeman effect, etc.) are frequent for stars. In order to correctly quantify the impact of the various physical processes that occur inside the atmospheres of such stars on the emergent spectra, one needs to test how the applied methods of spectral analysis allow one to fit the observations.  
\citet{2013A&A...551A..30B} determine the LTE abundances from lines of the first and second ions of
silicon in magnetic Bp and HgMn stars and use normal B stars of similar temperatures as a comparison sample. Both magnetic and non-magnetic stars show a discordance between the Si\ii\ and Si\iii\ based abundances, however, the effect is much smaller for the normal B-type stars, with the abundance difference Si\iii\ -- Si\ii\ = 0.3 to 0.8~dex for different stars. The departures from LTE are suggested as a source of these discrepancies.

In A to mid B-type stars, silicon can be observed in lines of the two ionisation stages: Si\ione\ and Si\ii, if the star's effective temperature ($\Teff$) is below approximately 11\,000~K, and Si\ii\ and  Si\iii\ for the hotter stars. This is suitable for constraining and checking atmospheric parameters. For different temperatures, either Si\ione\ or Si\ii\ is a minority species, for which the statistical equilibrium (SE) can easily deviate from thermodynamic equilibrium owing to deviations of the mean intensity of ionising radiation from the Planck function. Therefore, a correct evaluation of atmospheric parameters from the Si\ione /Si\ii\ or Si\ii /Si\iii\ ionisation equilibrium method can only be possible, when neglecting the LTE assumption.

\citet{2017PASJ...69...48S,2019PASJ...71...45S} registered weak emission lines of Si\ii\ in the red-region spectra of slowly rotating early B-type stars, and HD~160762 ($\iota$~Her, B3 IV) among them. The LTE calculations with classical hydrostatic model atmosphere can only yield an absorption line profile, but never the emission one. \citet{1972ApJ...175L..99M} were the first who showed that the emission line (N\iii\ 4634, 4640, 4641~\AA) in spectra of the main-sequence objects (O stars with $\Teff \ge$ 37\,000~K)
can be reproduced using classical hydrostatic model atmospheres, when taking the NLTE effects on the line formation into account. In $\iota$~Her, emission lines are observed for several chemical species \citep{2019PASJ...71...45S}. For C\ione\ and Ca\ii, their emission lines are well reproduced in the NLTE calculations of \citet{2016MNRAS.462.1123A} and \citet{2018MNRAS.477.3343S}, respectively.
%\citet{2001ApJ...546L.115S} Mn\ii\
%width of the emission lines is exactly the same as expected for absorption line in 

Non-LTE effects for Si\ione\ (and Si\ii, in some cases) are studied broadly for late-type and low-metallicity stars, using the model atoms treated by \citet{1976ApJS...30....1V,1978SoPh...56..263F,Shi_si_sun,2008ApJ...682.1376B,2012ApJ...755..176S,2013ApJ...764..115B}, and \citet{2017MNRAS.464..264A}. At the hottest end of spectral type sequence, early B and O, NLTE calculations for Si\ii, Si\iii, and Si\iv\ are performed 
%of in early B and O-type stars, with $\Teff \ge$ 15\,000~K, are studied 
with the model atoms constructed by \citet{1978ApJS...36..143K,1986MNRAS.222..719L}, and \citet{1990A&A...235..326B}. As for A to mid B-type stars, very few estimates of the NLTE abundances are based on theoretical results of \citet{2001A&A...373..998W}, who studied the Si\ii\ 3862, 4128-30, and 5041-56~\AA\ lines in a benchmark star Vega.
%Vernazza et al. (1976): 8 levels of Si I, sun, mostly for opacity calculations
%\citet{1978SoPh...56..263F}, Sun, Si  ii 1814 A, 8 levels of Si I + 8 levels of Si II
%\citet{2001A&A...373..998W} 75 si 1 and 40 si2, sun, vega
%\citet{Shi_si_sun} 132 terms of Si i, 41 terms of Si ii plus the Si iii ground state
%\citet{2008ApJ...682.1376B} sun, Si I 238 levels were reduced to 23
%\citet{2012ApJ...755..176S}: Sii+Siii, 296 fine structure levels, sun, 1.5D
%\citet{2013ApJ...764..115B} - Si I-Si II, NLTE Effects in J-band Silicon Lines

%\citet{1978ApJS...36..143K} Si II-IV, each ion separately; Si II 3853-67, 6347,71, 4128-30; 5041-56, total multiplets, Teff = 15000-35000 K
%\citet{1986MNRAS.222..719L} - Si II-IV, 47 levels
%\citet{1990A&A...235..326B} - $\Teff \ge$ 15000 K, Si II - 12 levels, Si\ii\ 4128-30, 5041-56, + Si III and IV

This study aimed to fill a gap in theoretical research of the line formation in stellar atmospheres and to develop the NLTE method for analysis of the Si\ione, Si\ii, and Si\iii\ lines in spectra of A to mid B-type stars. For a sample of nine unevolved A9 to B3-type stars with well determined atmospheric parameters, we investigated whether we can obtain for each star consistent abundances from different lines of silicon in different ionisation stages. A challenge for this study was understanding an origin of the Si\ii\ emission lines in $\iota$~Her.

The paper is organised as follows. The new model atom for Si\ione -\ii -\iii\ is presented in Sect.~\ref{sect:model_atom}. Our stellar sample, observational material, and the adopted atmospheric parameters are described briefly in Sect.~\ref{sect:obs}. The NLTE effects on atomic level populations and lines of Si\ione, Si\ii, and Si\iii\ are discussed in Sect.~\ref{sect:nlte}. Section~\ref{sect:mild} presents the abundance results for eight  stars with $\Teff \le$ 12800~K. Section~\ref{sect:iher} is devoted to our hottest star, $\iota$~Her, with its emission lines of Si\ii. Having convinced that the NLTE method works well for Si\ione -\ii -\iii\ through a range of A-B spectral types, we compute in Sect.~\ref{sect:dnlte} the NLTE abundance corrections for the silicon lines in the grid of model atmospheres. Our conclusions are summarised in Sect.~\ref{sect:Conclusions}.

%\section{The NLTE method for silicon}
\section{Model atom of Si\ione -Si\ii -Si\iii }\label{sect:model_atom}

{\it Energy levels.} For Si\ione\ we employ all the 451 energy levels below the ionisation threshold, $\chi_{\rm thr,1}$ = 65\,736.06~cm$^{-1}$ = 8.187~eV, provided by the NIST\footnote{https://physics.nist.gov/PhysRefData/ASD} database \citep{NIST19} based on the data of \citet{1983JPCRD..12..323M} and also the 268 high-excitation levels, which are predicted by R.~Kurucz\footnote{http://kurucz.harvard.edu/atoms/} in calculations of the Si\ione\ atomic structure and which are missing in NIST. The model atom includes the multiplet fine structure for all the terms up to Si\ione\ \eu{3p5s}{3}{P}{\circ}{}, with an excitation energy of \Eexc\ = 6.725~eV. 
%54\,528~cm$^{-1}$. 
For the \Eexc\ $>$ 7.96~eV 
%64\,200~cm$^{-1}$ 
levels, a combining procedure was applied, if the levels have a common parity and the energy separation does not exceed 0.037~eV (300~cm$^{-1}$). Thus, 95 levels in our model atom of Si\ione\ were made using the NIST data and 23 levels using the Kurucz's predictions.
%Finally, our model atom of Si\ione\ includes 118 levels. 
%Among them,  the remaining levels came from the predictions. 
%118 = 95 NIST + 23 Kurucz, fine structure up to 5s 3Po 

For Si\ii\ and Si\iii, the same sources of data were used, as for Si\ione. NIST provides 93 energy levels of Si\ii\ below the ionisation threshold, $\chi_{\rm thr,2}$ = 131\,838.14~cm$^{-1}$ = 16.346~eV. Taking into account the multiplet fine structure for the Si\ii\ ground (\eu{3p}{2}{P}{\circ}{}) and first excited (\eu{3p^2}{4}{P}{}{}) terms and neglecting it for all the other terms, we obtained 50 levels. We also implemented the 28 high-excitation levels predicted by R.~Kurucz, which were combined into six levels. Our model atom of Si\ii\ is complete up to principal quantum number $n$ = 9 and orbital quantum number $l$ = 7 and includes the terms with $l \le$ 4 for $n$ = 10 and 11.
%Si\ii\ 56  = 50 NIST + 6 Kurucz,  FS for 3p 2Po and 3p2 4P
%The model atom of Si\ii\ was complemented by 

For Si\iii, NIST provides 177 energy levels of the $3snl$ ($n \le 9$, $l \le 8$) and $3pnl$ ($n \le 4$, $l \le 2$) electronic configurations. Neglecting the multiplet fine structure for all the terms except the first excited one, \eu{3p}{3}{P}{\circ}{}, we obtained 54 levels in the model atom of Si\iii.

For $\Teff <$ 10\,000~K, the NLTE calculations were performed with a reduced model atom, which includes levels of Si\ione, Si\ii, and the ground state of Si\iii, because, in such atmospheres, a fraction of Si\iii\ is small
% minority species, %with $N$(Si\iii)/$N$(Si) $< 0.3$, 
and no Si\iii\ line can be detected in the $\Teff <$ 10\,000~K stars.
%the Si\iii\ 4552-4574\,\AA\ lines have non-measurable equivalent widths ($EW$s). 
For the higher temperatures, we used a model atom that includes four ionisation stages, that is Si\ione, Si\ii, Si\iii, and the ground state of Si\iv.

{\it Radiative bound-bound (b-b) transitions.} We do not take into account radiative transitions with either a wavelength of $\lambda >$ 300\,000\,\AA\ or oscillator strength of $f_{lu} < 10^{-6}$. Finally, radiative rates were computed for 1814, 475, and 327 allowed transitions of Si\ione, Si\ii, and Si\iii, respectively. Their $f_{lu}$-values were taken from calculations of R.~Kurucz. In case, if either lower or/and upper level of the transition is combined, the average-''multiplet'' oscillator strength was calculated as $f_{lu}$ = $\Sigma g_{i}f_{ij} /g_{l}$, where summing is performed for all the levels $i$ and $j$, which constitute the levels $l$ and $u$, respectively; $g_{i}$ and $g_{l}$ are the statistical weights of the levels $i$ and $l$. 

{\it Radiative bound-free (b-f) transitions.} 
For 55 lowest levels of Si\ione\ with angular momentum $L \le 3$, up to \eu{3p7s}{3}{P}{\circ}{} (the threshold wavelength is $\lambda_{\rm thr}$ = 24\,000\,\AA), 41 levels of Si\ii\ ($L \le 4$, up to \eu{9p}{2}{P}{\circ}{} with $\lambda_{\rm thr}$ = 14\,690\,\AA), and 40 levels of Si\iii\ up to \eu{7d}{3}{D}{}{} ($\lambda_{\rm thr}$ = 4742\,\AA), we rely on the photoionisation cross sections calculated within the Opacity Project \citep[OP,][]{1987JPhB...20.6363S} and accessible in the TOPbase\footnote{http://cdsweb.u-strasbg.fr/cgi-bin/topbase/} database \citep{1993A&A...275L...5C}. For the remaining levels in the model atom, the hydrogenic approximation was used with effective principal quantum number $n_{eff}$ instead of $n$.

We note that the photoionisation cross sections computed by \citet{2000ApJS..126..537N} and \citet{1995ApJS..101..423N} for Si\ione\ and Si\ii, respectively, as available in the NORAD\footnote{https://norad.astronomy.osu.edu/} database, are very similar to the TOPbase's ones for photon energies close to the ionisation threshold within approximately 0.5~Ryd, and they are systematically smaller at the higher photon energies, by 0.1 to 0.5~dex for different levels.
 \citet{2011CaJPh..89.1119S} calculated for the first time the photoionisation cross sections for the fine-splitting levels of the Si\ii\ \eu{3s^2 3p}{2}{P}{\circ}{}, \eu{3s 3p^2}{4}{P}{}{}, and \eu{3s 3p^2}{2}{D}{}{} terms. Since these data are not available in a tabular form, they cannot be checked in our NLTE calculations. However, we note that in the stellar parameter range, with which this study concerns, photoionisation of the \eu{3s^2 3p}{2}{P}{\circ}{} and \eu{3s 3p^2}{4}{P}{}{} terms affects only weakly the SE of Si\ii. For Si\ii\ \eu{3s 3p^2}{2}{D}{}{}, the photoionisation cross sections in Figures~6 and 7 of \citet{2011CaJPh..89.1119S} are smaller than the TOPbase ones for photon energies below 1.2~Rydberg and resemble predictions of \citet{1995ApJS..101..423N}. 

%  by \citet{2000ApJS..126..537N} and  \citet{1995ApJS..101..423N} are carried out in the close coupling approximation using the R-matrix method.
% {1993JPhB...26.1109N}

{\it Collisional transitions.} For Si\ione\ the electron-impact excitations are not yet known with sufficient accuracy, and our calculations of collisional
rates rely on theoretical approximations. We used the formula of \citet{Reg1962} for the allowed transitions and assumed that the effective collision strength $\Upsilon$ = 1 for the forbidden transitions. 
For 414 transitions between the terms up to \eu{3s3p3d}{2}{P}{\circ}{} in Si\ii\ (28\,\%\ of the total number of collisional transitions in Si\ii) 
%\Eexc\ = 126\,235~cm$^{-1}$
 and 210 transitions between the terms up to \eu{3p3d}{1}{F}{\circ}{} in Si\iii\ (15\,\%\ of the total number of collisional transitions in Si\iii),
 %\Eexc\ = 235\,413~cm$^{-1}$),
 we employed the data from the R-matrix calculations of \citet{2014MNRAS.442..388A} and \citet{2014A&A...572A.115F}, respectively. 
The same formulas as for Si\ione\ were applied for the remaining transitions in Si\ii\ and Si\iii.  Ionisation by electronic collisions was calculated everywhere from the \citet{1962amp..conf..375S} formula with the hydrogenic photoionization cross section at threshold.

%Si II collisions \citet{2014MNRAS.442..388A} 3s.3p.(3P*).3d 2P*
%Si\iii\ \citet{2014A&A...572A.115F}
%forbidden: Upsilon = 1 for Si I-II-III
The effects of uncertainties in atomic data on the NLTE results are discussed in Sect.~\ref{sect:iher}, using the silicon emission and absorption lines in $\iota$~Her.

\begin{table*}
	\centering
	\caption{Atmospheric parameters of the sample stars and mean NLTE (N) and LTE (L) abundances from lines of different ionisation stages of silicon.}
	\label{tab:stars_param}
	\begin{tabular}{lrcrccrrccrlcccr} % 
		\hline\hline \noalign{\smallskip}
\multicolumn{1}{c}{HD} & $\Teff$ & $\logg$ & [Fe/H] & $\xi_t$ & Ref & & \multicolumn{2}{c}{Si\ione} & & \multicolumn{2}{c}{Si\ii} & & \multicolumn{2}{c}{Si\iii} & [Si/H] \\
\cline{8-9}
\cline{11-12}
\cline{14-15}
         & [K]  &      &      &  &   &  &  $N_l$ & $\eps{}$   & &  $N_l$ & \multicolumn{1}{c}{$\eps{}$} & & $N_l$ & $\eps{}$ &  \\
\noalign{\smallskip}\hline \noalign{\smallskip}
 \ 32115 & 7250 & 4.20 & 0.00 & 2.3 & F11 & N & 16 & 7.58(0.17) & &  4 & 7.65(0.08) & & &  &0.10~ \\
         &      &      &      &     &     & L &    & 7.61(0.17) & &    & 7.82(0.24) & & &  &\\
 \ 73666 & 9380 & 3.78 & 0.10 & 1.8 & F07 & N &  1 & 7.54       & & 13 & 7.68(0.17) & & &  &0.10~ \\
(40 Cnc) &      &      &      &     &     & L &    & 7.16       & &    & 7.84(0.26) & & &  &\\
172167  & 9550 & 3.95 & $-$0.50 & 1.8 & C93 & N &    &            & &  8 & 6.92(0.15) & & &  &$-$0.59~ \\
(Vega)    &      &      &       &   &     & L &    &            & &    & 7.08(0.17) & & &  &\\
 \ 72660 & 9700 & 4.10 & 0.40 & 1.8 & S16 & N &  6 & 7.85(0.04) & & 14 & 7.82(0.11) & & &  & 0.33~ \\
         &      &      &      &     &     & L &    & 7.59(0.09) & &    & 7.94(0.23) & & &  &\\
145788   & 9750 & 3.70 & 0.46 & 1.3 & F09 & N &    &            & &  7 & 7.63(0.12) & & &  & 0.12~ \\
         &      &      &      &     &     & L &    &            & &    & 7.90(0.19) & & &  &\\
 \ 48915 & 9850 & 4.30 & 0.40 & 1.8 & H93 & N &  1 &  7.59      & & 11 & 7.72(0.11) & & &  & 0.21$^1$ \\
(Sirius) &      &      &      &     &     & L &    &  7.26      & &    & 7.84(0.18) & & &  &\\
209459  & 10400 & 3.55 & 0.00 & 0.5 & F09 & N &  1 & 7.49       & & 17 & 7.50(0.13) & & &  &$-$0.01~ \\
(21 Peg)  &       &      &      &   &     & L &    & 7.08       & &    & 7.62(0.29) & & &  &\\
 \ 17081 & 12800 & 3.75 & 0.00 & 1.0& F09 & N &    &            & & 19 & 7.75(0.14) & & 2 & 7.72(0.04) & 0.23~ \\
($\pi$ Cet) &    &      &      &    &     & L &    &            & &    & 7.59(0.25) & &   & 7.82(0.00) & \\
160762  & 17500 & 3.80 & 0.02 & 1.0 & N12 & N &    &            & & 13 & 7.74(0.25)$^2$ & & 4 & 7.54(0.07) & 0.03$^3$ \\
($\iota$ Her) &    &      &      &  &     & L &    &            & &    & 7.07(0.41)$^2$ & &   & 7.79(0.05) & \\
%  & \multicolumn{5}{l}{emission lines}    & NLTE &    &            & &  6 & 8.10(0.26) & &   &            \\   
\noalign{\smallskip}\hline \noalign{\smallskip}
\multicolumn{16}{l}{{\bf Notes.} The numbers in parentheses are the dispersions in the single line measurements around the mean. $\xi_t$ is in \kms. } \\
\multicolumn{16}{l}{ \ \ $^1$ based on Si\ii\ lines; $^2$ Si\ii\ 6347, 6371\,\AA\ and emission lines are not accounted in the mean; $^3$ based on Si\iii\ lines. } \\
\multicolumn{16}{l}{{\bf Ref:} C93 = \citet{1993ASPC...44..496C}, F07, F09, F11 = \citet{2007AA...476..911F,2009AA...503..945F,2011MNRAS.417..495F}, } \\
\multicolumn{16}{l}{H93 = \citet{1993AA...276..142H}, N12 = \citet{2012A&A...539A.143N}, S16 = \citet{sitnova_ti}.} 
	\end{tabular}
\end{table*}

\section{Stellar sample, observations, atmospheric parameters}\label{sect:obs}

As a test and first application of the model atom, silicon lines in nine objects were analysed.
This research continues a series of the NLTE studies of chemical species in A-B type stars, namely, O\ione\ \citep{sitnova_o}, C\ione -\ii\ \citep{2016MNRAS.462.1123A}, Ti\ione -\ii\ \citep{sitnova_ti}, Mg\ione -\ii\ \citep{2018ApJ...866..153A}, Ca\ione -\ii\ \citep{2018MNRAS.477.3343S}. Therefore, we take the same stellar sample with their atmospheric parameters and 
 the same spectral observations. The selected stars, their effective temperatures, surface gravities ($\logg$), metallicities ([Fe/H]), microturbulent velocities ($\xi_t$), and the sources of these data are listed in Table~\ref{tab:stars_param}. 
 
 \citet{2016MNRAS.462.1123A} described in detail the programme stars. We briefly summarise. Only sharp-lined stars, with rotational velocities of $V \sin i \precsim 20$~\kms, were included in our sample, in order to do spectral analysis at the highest precision.
 
The stars 21~Peg and $\iota$~Her are chemically normal single stars. 

Each of HD~32115, HD~73666, and $\pi$~Cet is a primary component of single line spectroscopic binary (SB1), with negligible flux coming from the secondary star. Therefore, it is safe to analyse these stars ignoring the presence of their secondaries. For none of them, a chemical peculiarity was reported. HD~73666 is a Blue Straggler and a member of the Praesepe cluster.
  
HD~145788 reveals an overabundance of almost all metals, according to \citet{2009AA...503..945F}. They conclude that its element abundance pattern 'could be explained if HD~145788 was formed in a region of the sky with a metallicity higher than the solar region'.

Sirius is an astrometric visual binary system composed of a main-sequence A1V star and a DA white dwarf. The primary component is classified as a metallic-line (Am) star.

HD~72660 was classified by \citet{2016MNRAS.456.3318G} as a transition object between an HgMn star and an Am star.

Vega is a rapidly rotating star seen pole-on and is also classified as a mild $\lambda$~Bootis-type star. Similarly to our previous studies, we ignore the non-spherical effects and analyse Vega's flux spectrum using the average temperature and gravity.

%{\bf The star $\pi$~Cet is a SB1 star with $V\sin i \sim$ 20~\kms, however its spectrum is not visibly contaminated by the companion. 

 We refer to the original papers and also \citet{2016MNRAS.462.1123A,2018ApJ...866..153A}, and \citet{sitnova_ti} for a description of the methods of atmospheric parameter determinations. 
%The star pi Cet is a known binary with a period of about 7.5 years (Lacy et al. 1997) and is a Herbig AeBe star (Malfait et al. 1998). The Herbig classification comes from a detected infrared excess at wavelengths longer than 10 μm. The two spectra obtained with ESPaDOnS show variability in the line profiles, small emission-like features close to the core of Halpha, and emission features at the position of C i 8335 and 9405 in the near infrared. The pre-main sequence status of this star, which is very likely responsible for these emissions, The variation in the line profile within one day excludes the possibility that the observed changes are caused by the companion
% i Her is a single bright star

For the seven of nine stars, observational material for the visual spectral range was obtained with a spectral resolving power of $R = \lambda/\Delta\lambda >$ 60\,000 and a signal-to-noise ratio of S/N $>$ 200, using the ESPaDOnS instrument of the Canada-France-Hawaii Telescope (CHFT).
% HD~73666 SB1, ESPaDOnS, 65 000 /660; F07
% pi Cet SB1, ESPaDOnS instrument of the CHFT, F09
% HD 72660, ESPaDOnS K18
% HD 145788, échelle spectrograph HARPS instrument attached at the 3.6-m ESO La Silla telescope,  115 000 / 200; F09
% , namely, HD~17081, HD~32115, HD~72660, HD~73666, HD~160762, and HD~209459, Sirius
%is described in detail by \citet[][see their Table~3]{2018MNRAS.477.3343S}. 
%We briefly summarise: 
Vega was observed by A.~Korn using the spectrograph FOCES ($R \simeq$ 40\,000, S/N $>$ 750) at the 2.2~m telescope of the Calar Alto Observatory. HD~145788 was observed with the {\'{e}}chelle spectrograph HARPS instrument ($R \simeq$ 115 000, S/N $\simeq$ 200) attached at the 3.6-m ESO La Silla telescope.
For HD~72660, we use also the UV spectrum kindly provided by J.~Landstreet. The observations and spectrum reduction are described by \citet{2016MNRAS.456.3318G}.
%UV spectrum of HD 72660: Golriz, Landstreet 2016
%The atmospheric parameters are listed in Table~\ref{tab:stars_param}. 

%\section{Determination of NLTE abundances}
\section{Non-LTE calculations}\label{sect:nlte}

\subsection{Codes, model atmospheres, list of investigated lines}

The coupled radiative transfer and SE equations were solved with a modified version of the {\sc DETAIL} code \citep{detail,2011JPhCS.328a2015P}. As verified by \citet{2011JPhCS.328a2015P}, a hybrid method combining LTE atmospheres and NLTE line formation is able to reproduce observations for effective temperatures between 15\,000 and 35\,000~K. This is all the more true for cooler atmospheres. 
%This study concerns with the 7250 $\le \Teff \le$ 17\,500~K range and nine A-B type unevolved stars. 
For consistency with our earlier studies, for each star we used exactly the same model atmosphere. 
Classical plane-parallel and LTE model atmospheres were calculated with the code \textsc{LLmodels} \citep{2004AA...428..993S}. For Sirius, its model atmosphere was taken from R.~Kurucz website\footnote{http://kurucz.harvard.edu/stars/sirius/ap04t9850g43k0he05y.dat}.

Linelists of \citet{2009AA...503..945F,2011MNRAS.417..495F}, and \citet{2019PASJ...71...45S} were used to select the silicon lines for the non-LTE analysis. A sample of lines observed in stellar spectra changes substantially, when moving from our coolest to the hottest star. Therefore, the used lines together with their atomic parameters are listed in five separate Tables: Si\ione\ and Si\ii\ lines in HD~32115 (Table~\ref{tab:sun_hd32115}), Si\ione\ UV lines (Table~\ref{tab:hd72660}), Si\ione\ and Si\ii\ lines in the visible spectra of eight stars with $\Teff >$ 9000~K (Table~\ref{tab:abund_stars}), Si\iii\ lines (Table~\ref{tab:abund_si3}), and Si\ii\ emission lines in $\iota$~Her (Table~\ref{tab:emission_si2}). For lines of Si\iii, the quadratic Stark effect broadening was taken into account using approximate formula of 
\citet{1971Obs....91..139C}.

In this research, analysis of observed spectra is based on line profile fitting. The synthetic spectra were computed with the {\sc Synth}V\_NLTE code \citep{2019ASPC}, which implements the pre-computed departure coefficients from the {\sc DETAIL} code. We note that {\sc Synth}V\_NLTE treats a contribution of the overlapping hydrogen Balmer lines to backgroung opacity using the occupation probability theory, as developed by \citet{1988ApJ...331..794H}, \citet{1994A&A...282..151H}, and \citet{1999ApJ...526..451N}. This is important for calculations of Si\ione\ 3905\,\AA\ and Si\ii\ 3853, 3856, 3862\,\AA.
The best fit to the observed spectrum was obtained automatically using the {\sc IDL binmag} code by O. Kochukhov\footnote{http://www.astro.uu.se/$\sim$oleg/binmag.html}.  The line list and atomic data for the synthetic spectra calculations were taken from the VALD database \citep{2015PhyS...90e4005R}.

\begin{figure}
	% To include a figure from a file named example.*
	% Allowable file formats are eps or ps if compiling using latex
	% or pdf, png, jpg if compiling using pdflatex
	\includegraphics[width=\columnwidth]{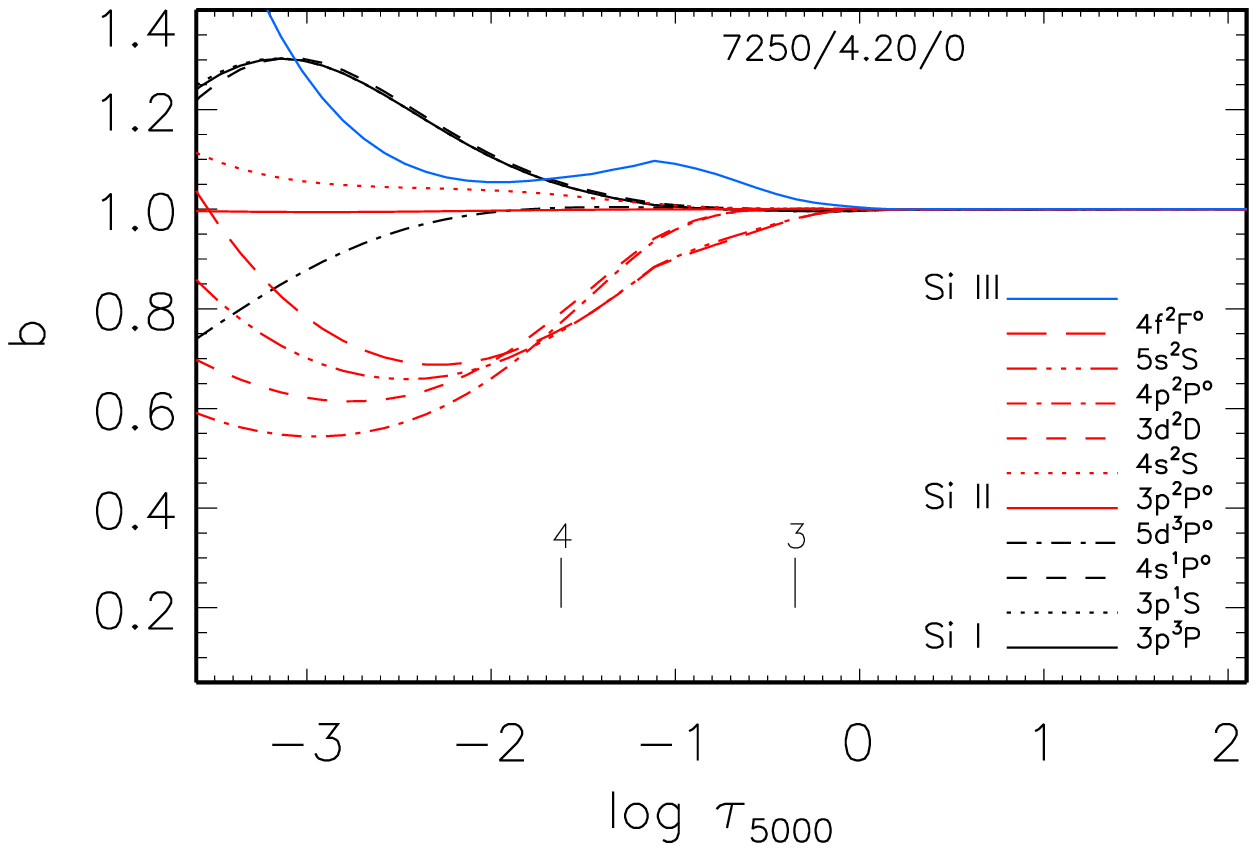}
	
	\vspace{-5mm}
	\includegraphics[width=\columnwidth]{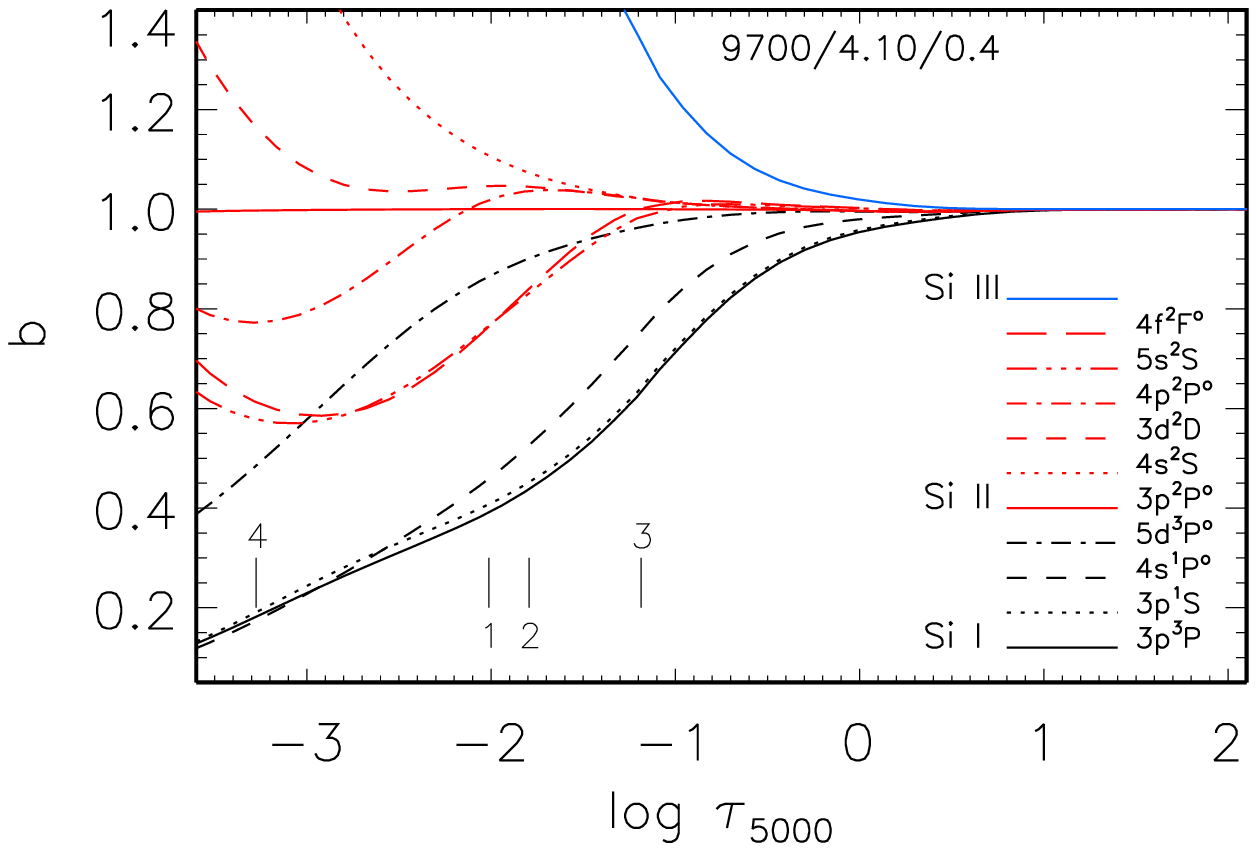}

	\vspace{-5mm}
	\includegraphics[width=\columnwidth]{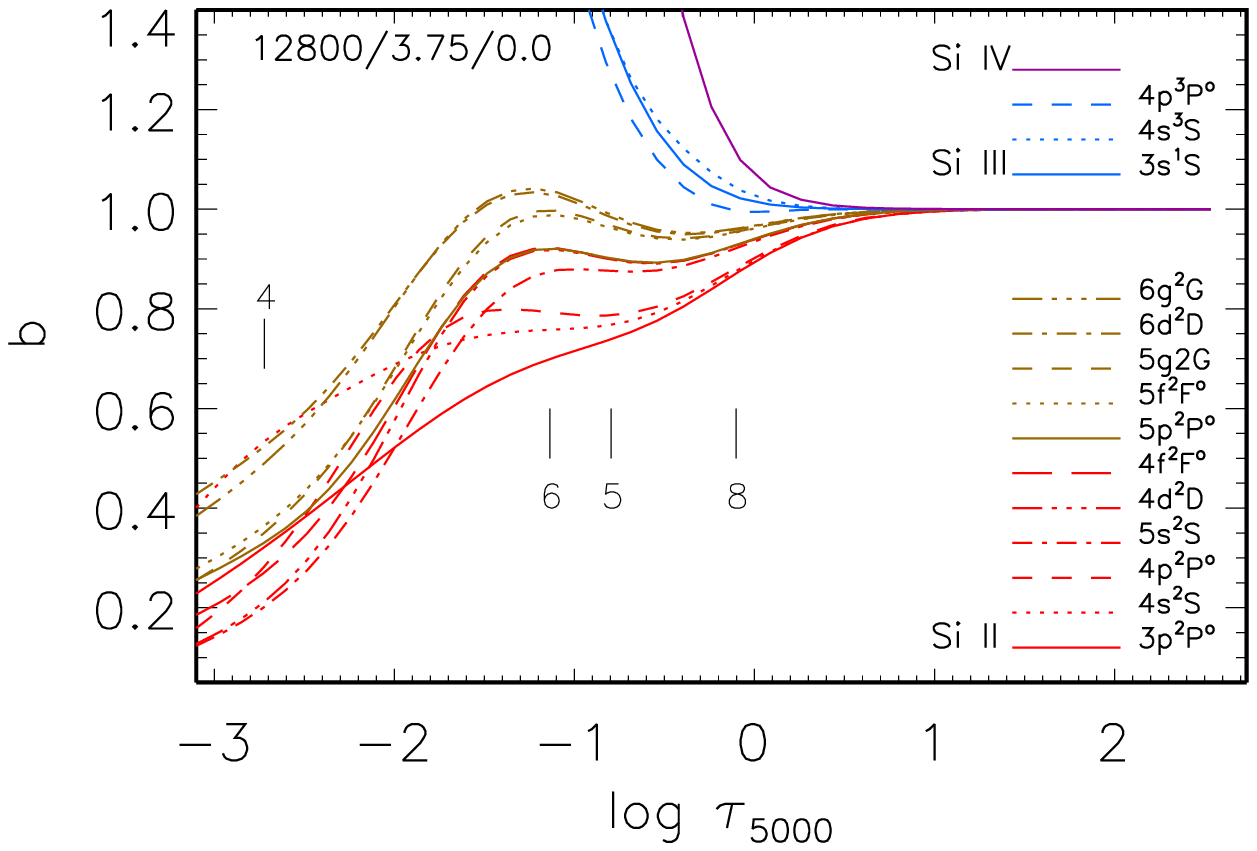}

	\vspace{-5mm}
	\includegraphics[width=\columnwidth]{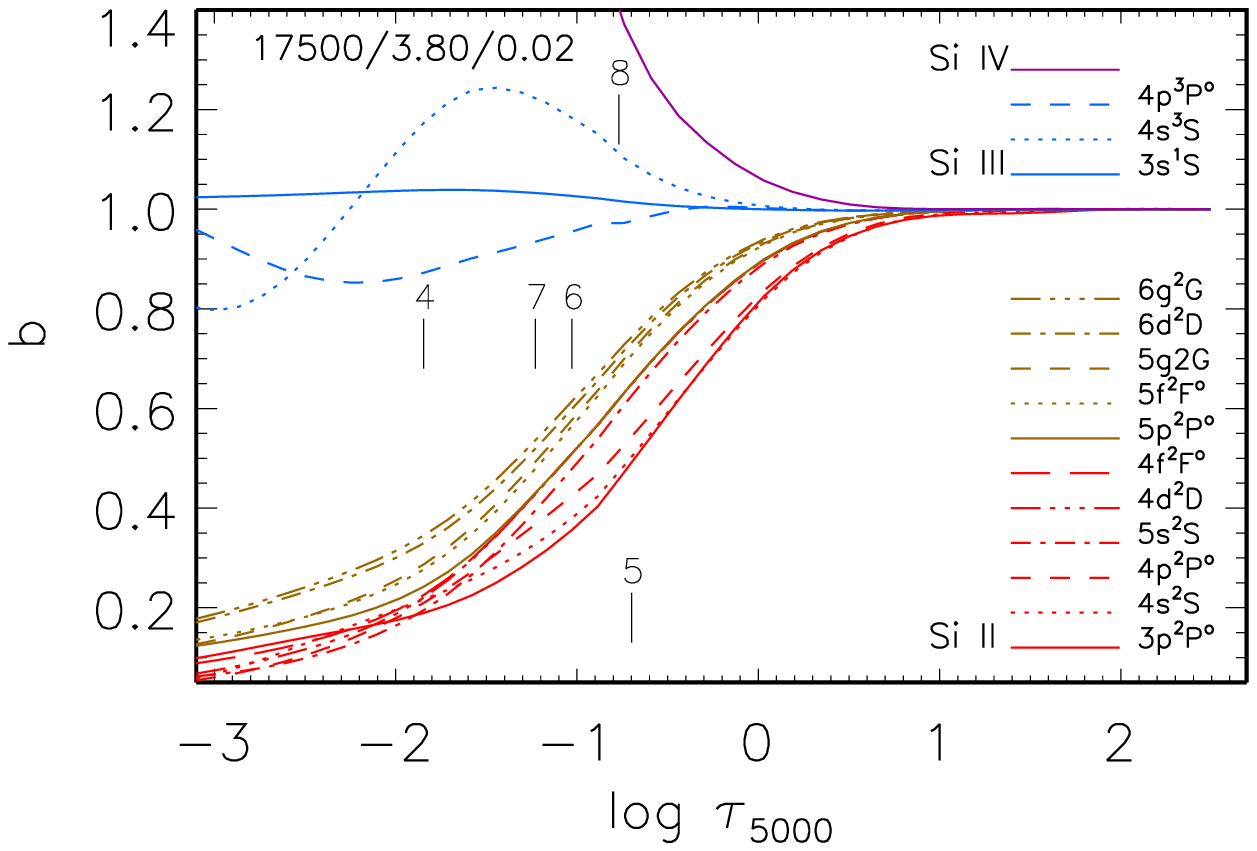}

	\vspace{-5mm}
    \caption{Departure coefficients, $b$, for the levels of Si\ione\ (black curves, two top panels), Si\ii\ (red and brown curves), Si\iii\ (blue curves), and Si\iv\ (lilac curves) as a function of $\log \tau_{5000}$ in the model atmospheres 7250/4.20/0, 9700/4.10/0.4, 12800/3.75/0.0, and 17500/3.80/0.02. The selected levels are quoted in the right part of each panel. Tick marks indicate the locations of 
line center optical depth unity for the following lines: Si\ione\ 1666 (1) and 3905\,\AA\ (2), Si\ii\ 5957 (3), 6347 (4), 6239 (5), 7848 (6), and 9412\,\AA\ (7) and Si\iii\ 4567\,\AA\ (8). }
    \label{fig:bfactors}
\end{figure}

\subsection{Statistical equilibrium of silicon depending on effective temperature}
%\subsection{Departures from LTE in atomic level populations}

For the energy levels, which are important for understanding a formation of the silicon lines, which are observed in our sample stars, Fig.~\ref{fig:bfactors} displays the departure coefficients, ${\rm b = n_{NLTE}/n_{LTE}}$, in four model atmospheres with different $\Teff$.  Here,
${\rm n_{NLTE}}$ and ${\rm n_{LTE}}$ are the statistical equilibrium and thermal (Saha-Boltzmann) number densities, respectively. 
In our coolest atmosphere ($\Teff$ / $\logg$ = 7250~K / 4.20), silicon is strongly ionised, with $N$(Si\ii)/$N$(Si\ione) $> 30$. For Si\ione, the departures from LTE take place above continuum optical depth $\log \tau_{5000} \simeq -1$.
Superthermal radiation of a non-local origin below the thresholds of the Si\ione\ levels, such as \eu{3p^2}{1}{D}{}{} ($\lambda_{\rm thr}$ = 1682\,\AA), \eu{3p4s}{1}{P}{\circ}{} ($\lambda_{\rm thr}$ = 4039\,\AA), and \eu{3p3d}{1}{D}{\circ}{} ($\lambda_{\rm thr}$ = 5435\,\AA), tends to overionise Si\ione. However, bound-bound transitions from many levels close to the ionisation limit down to the lower levels 
%cause a photon suction. 
can siphon an efficient flow of electrons downward. It
 increases the populations of the ground state, \eu{3p^2}{3}{P}{}{}, and low-lying levels, such as \eu{3p^2}{1}{S}{}{} and \eu{3p4s}{1}{P}{\circ}{}, above their thermodynamic equilibrium (TE) values, but depopulates the high-excitation (\Eexc\ $\gtrsim$ 6~eV) levels, such as \eu{3p5d}{3}{P}{\circ}{} (in the top panel of Fig.~\ref{fig:bfactors}, it is quoted as \eu{5d}{3}{P}{\circ}{} for brevity). The levels above \eu{3d}{2}{D}{}{} in Si\ii\ are underpopulated (${\rm b} < 1$) by photon losses in the transitions to the lower levels as soon as the line center optical depth drops below 1. Two transitions from the ground state to \eu{4s}{2}{S}{}{} ($\lambda$ = 1526, 1533\,\AA) are close to be in detailed balance throughout the atmosphere. This explains why population of \eu{4s}{2}{S}{}{} is close to the TE value.

In the hotter model atmosphere (9700/4.10), no processes in Si\ione\ can compete with overionisation of the low-lying levels, resulting in strong depopulation of all the Si\ione\ levels. 
Si\ii\ remains to be a majority species, and its ground state keeps the TE population throughout the atmosphere. Below log~$\tau_{5000} \simeq -1$ the most important transitions of Si\ii\ are in detailed balance and the departure coefficients of the excited levels are close to 1. In the higher atmospheric layers, strong pumping transitions \eu{3p}{2}{P}{\circ}{} -- \eu{4s}{2}{S}{}{} ($\lambda$ = 1526, 1533\,\AA) and \eu{3p}{2}{P}{\circ}{} -- \eu{3d}{2}{D}{}{} ($\lambda$ = 1260, 1264\,\AA) 
produce enhanced excitation of the upper levels, while photon losses in the transitions to low-lying levels depopulate \eu{4p}{2}{P}{\circ}{} and the levels above. 

Si\ione\ is strongly overionised in the two hottest models, 12\,800/3.75 and 17\,500/3.80, which represent the atmospheres of $\pi$~Cet and $\iota$~Her. Since no lines of Si\ione\ can be measured in these stars, levels of Si\ione\ are not shown in Fig.~\ref{fig:bfactors}. In the 12\,800/3.75 model, Si\ii\ and Si\iii\ have comparable number densities in the line-formation layers, above log~$\tau_{5000} = 0$. Superthermal radiation below the thresholds of the Si\ii\ levels, such as \eu{4s}{2}{S}{}{} ($\lambda_{\rm thr}$ = 1507\,\AA), \eu{3d}{2}{D}{}{} ($\lambda_{\rm thr}$ = 1905\,\AA), and \eu{3p^2}{2}{D}{}{} ($\lambda_{\rm thr}$ = 1306\,\AA), leads to an overionisation of Si\ii, but enhanced populations of the Si\iii\ levels.
% compared with the TE populations.

In the 17\,500/3.80 model, Si\ii\ is subject to strong overionisation, while Si\iii\ is a majority species. Its ground state keeps close to the TE population throughout the atmosphere. Pumping UV transitions from the low-excitation levels produce enhanced excitation of the lower level of the Si\iii\ \eu{4s}{3}{S}{}{} -- \eu{4p}{3}{P}{\circ}{} transition (triplet lines at 4552, 4567, and 4574\,\AA) above log~$\tau_{5000} \simeq 0$, while the upper level is depopulated via spontaneous transitions.

\begin{table}
	\centering
	\caption{NLTE (N) and LTE (L) abundances, $\eps{}$, from lines of Si\ione\ and Si\ii\ in the Sun and HD~32115.}
	\label{tab:sun_hd32115}
	\begin{tabular}{lcrrcc} % 
		\hline\hline \noalign{\smallskip}
Transition & \multicolumn{1}{c}{ $\lambda$} & log $gf$ &  & Sun & HD \\ 
           & \multicolumn{1}{c}{ [\AA]}     &          &  &     & 32115 \\
\noalign{\smallskip}\hline \noalign{\smallskip}
\multicolumn{1}{l}{Si\ione} & & & & & \\
 \eu{4s}{3}{P}{\circ}{} - \eu{5p}{3}{P}{}{} & 5690.42 & $-1$.87$^1$ & N & 7.64 &   7.55 \\
 \Eexc\ =  4.93~eV                          &         &           &  L & 7.66 &   7.57 \\
 \eu{4s}{1}{P}{\circ}{} - \eu{5p}{1}{S}{}{} & 5772.15 & $-1$.75$^1$ & N & 7.65 &   7.60 \\
 \Eexc\ =  5.08~eV                          &         &           &  L & 7.67 &   7.61 \\
 \eu{4s}{1}{P}{\circ}{} - \eu{5p}{1}{D}{}{} & 5948.54 & $-1$.23$^1$ & N & 7.64 &   7.64 \\
 \Eexc\ =  5.08~eV                          &         &           &  L & 7.69 &   7.67 \\
 \eu{3p^3}{3}{D}{\circ}{} - \eu{5f}{3}{D}{}{} & 6142.48 & $-1$.30$^2$ &  N & 7.37 &   7.40 \\
 \Eexc\ =  5.62~eV                          &         &           &  L & 7.39 &   7.41 \\
 \eu{3p^3}{3}{D}{\circ}{} - \eu{5f}{3}{G}{}{} & 6155.13 & $-0$.76$^2$ &  N & 7.46 &   7.42 \\
 \Eexc\ =  5.62~eV                          &         &           &   L & 7.50 &   7.44 \\
 \eu{3p^3}{3}{D}{\circ}{} - \eu{5f}{3}{F}{}{} & 6237.32 & $-0$.98$^2$ &  N & 7.40 &   7.39 \\
 \Eexc\ =  5.61~eV                          &         &           &   L & 7.42 &   7.41 \\
 \eu{3d}{1}{D}{\circ}{} - $6f$\scriptsize{2[7/2]} & 6414.98 & $-1$.04$^2$ &  N & 7.51 &   7.51 \\
 \Eexc\ =  5.87~eV                          &         &           &   L & 7.53 &   7.53 \\
 \eu{4p}{1}{P}{}{} - \eu{6d}{1}{D}{\circ}{} & 6721.85 & $-0$.94$^3$ &  N & 7.36 &   7.31 \\
 \Eexc\ =  5.86~eV                          &         &           &   L & 7.38 &   7.33 \\
 \eu{4p}{3}{D}{}{} - \eu{6d}{3}{P}{\circ}{} & 6741.63 & $-1$.75$^2$ &  N & 7.68 &   7.80 \\
 \Eexc\ =  5.98~eV                          &         &           &   L & 7.70 &   7.83 \\
 \eu{4p}{1}{P}{}{} - $7s$\scriptsize{(3/2,1/2)}$^\circ$ & 6848.58 & $-1$.53$^2$ &  N & 7.42 &   7.32 \\
 \Eexc\ =  5.86~eV                          &         &           &   L & 7.43 &   7.34 \\
 \eu{3d}{1}{D}{\circ}{} - $5f$\scriptsize{2[7/2]}     & 7034.90 & $-0$.88$^2$ &  N & 7.64 &   7.57 \\
 \Eexc\ =  5.87~eV                          &         &           &   L & 7.68 &   7.61 \\
 \eu{4p}{3}{D}{}{} - $7s$\scriptsize{(3/2,1/2)}$^\circ$ & 7373.00 & $-1$.18$^2$ &  N & 7.43 &   7.37 \\
 \Eexc\ =  5.98~eV                          &         &           &   L & 7.43 &   7.39 \\
 \eu{3p^3}{3}{D}{\circ}{} - \eu{4f}{3}{F}{}{} & 7405.77 & $-0$.82$^2$ &  N & 7.61 &   7.65 \\
 \Eexc\ =  5.61~eV                          &         &           &   L & 7.67 &   7.70 \\
 \eu{4p}{1}{P}{}{} - \eu{5d}{1}{D}{\circ}{} & 7680.27 & $-0$.69$^1$ &  N & 7.65 &   7.57 \\
 \Eexc\ =  5.86~eV                          &         &           &   L & 7.71 &   7.61 \\
 \eu{3d}{3}{F}{\circ}{} - $6f$\scriptsize{2[5/2]}     & 7849.97 & $-0$.71$^2$ &  N & 7.50 &   7.50 \\
 \Eexc\ =  6.19~eV                          &         &           &   L & 7.53 &   7.53 \\
 \eu{4p}{3}{D}{}{} - \eu{5d}{3}{F}{\circ}{} & 7944.00 & $-0$.31$^1$ &  N & 7.62 &   7.56 \\
 \Eexc\ =  5.98~eV                          &         &           &  L  & 7.72 &   7.64 \\
\multicolumn{1}{l}{Si\ii} & &       &      &        \\
 \eu{4p}{2}{P}{\circ}{} - \eu{4d}{2}{D}{}{} & 5055.98 &  0.52$^4$ &  N &      &   7.68 \\
 \Eexc\ = 10.07~eV                          &         &           &   L &      &   7.68 \\   
 \eu{4p}{2}{P}{\circ}{} - \eu{5s}{2}{S}{}{} & 5978.93 &  0.08$^5$ &  N &      &   7.58 \\ 
 \Eexc\ = 10.07~eV                          &         &           &   L &      &   7.58 \\    
 \eu{4s}{2}{S}{}{} - \eu{4p}{2}{P}{\circ}{} & 6347.11 &  0.15$^4$ &  N & 7.62 &   7.75 \\
  \Eexc\ = 8.12~eV                          &         &           &   L & 7.72 &   8.04 \\
 & 6371.37 & $-0$.08$^4$ &  N & 7.50 &   7.64 \\
          &       &           &   L & 7.58 &   7.86 \\
\noalign{\smallskip}\hline \noalign{\smallskip}
\multicolumn{6}{l}{{\bf Notes.} $^1$ \citet{1973A&A....26..471G}, $^2$ VALD, $^3$ \citet{1969atp..book.....W},  } \\
\multicolumn{6}{l}{ \ \ \ \ $^4$ \citet{2001JQSRT..69..535M}, $^5$ \citet{1995PhyS...52..628B}.}		
	\end{tabular}
\end{table}

\begin{table}
	\centering
	\caption{NLTE and LTE abundances, $\eps{}$, from the UV lines of Si\ione\ and Si\ii\ in HD~72660.}
	\label{tab:hd72660}
	\begin{tabular}{lccrcc} % 
		\hline\hline \noalign{\smallskip}
Transition & \multicolumn{1}{c}{ $\lambda$} & \Eexc & log $gf$ & NLTE & LTE \\ 
           & \multicolumn{1}{c}{ [\AA]}     & [eV]  &          &      & \\
\noalign{\smallskip}\hline \noalign{\smallskip}
\multicolumn{6}{l}{Si\ione\ } \\
% \eu{3p}{3}{P}{}{2} - \eu{5d}{1}{D}{\circ}{2} & 1664.51 & 0.03 & $-1$.80 &  7.86 &   7.66   \\
% \eu{3p}{3}{P}{}{0} - \eu{5d}{3}{P}{\circ}{1} & 1666.38 &  0.00 & -1.66 &  7.84 &   7.54  \\
% \eu{3p}{3}{P}{}{1} - \eu{6s}{1}{P}{\circ}{2} & 1682.67 &  0.01 & -1.85 &  7.91 &   7.64  \\ 
% \eu{3p}{3}{P}{}{1} - \eu{6s}{3}{P}{\circ}{1} & 1689.29 &  0.01 & -1.83 &  7.88 &   7.66  \\ 
% \eu{3p}{3}{P}{}{2} - \eu{4d}{3}{D}{\circ}{1} & 1699.72 &  0.03 & -2.63 &  7.82 &   7.58   \\
 \eu{3p}{3}{P}{}{} - \eu{5d}{1}{D}{\circ}{} & 1664.51 & 0.03 & $-1$.80 &  7.86 &   7.66   \\
 \eu{3p}{3}{P}{}{} - \eu{5d}{3}{P}{\circ}{} & 1666.38 &  0.00 & $-1$.66 &  7.84 &   7.54  \\
 \eu{3p}{3}{P}{}{} - \eu{6s}{1}{P}{\circ}{} & 1682.67 &  0.01 & $-1$.85 &  7.91 &   7.64  \\ 
 \eu{3p}{3}{P}{}{} - \eu{6s}{3}{P}{\circ}{} & 1689.29 &  0.01 & $-1$.83 &  7.88 &   7.66  \\ 
 \eu{3p}{3}{P}{}{} - \eu{4d}{3}{D}{\circ}{} & 1699.72 &  0.03 & $-$2.63 &  7.82 &   7.58   \\
\multicolumn{6}{l}{Si\ii } \\
% \eu{3p^2}{2}{D}{}{3/2} - \eu{5f}{2}{F}{\circ}{5/2} & 1710.84 &  6.86 & -0.57 & 7.97 &   7.97 \\
 \eu{3p^2}{2}{D}{}{} - \eu{5f}{2}{F}{\circ}{} & 1710.84 &  6.86 & $-0$.57 & 7.97 &   7.97 \\
\noalign{\smallskip}\hline \noalign{\smallskip}
\multicolumn{6}{l}{{\bf Notes.} $gf$-values are from \citet{1987ApJ...322..573S} for Si\ione,} \\
\multicolumn{6}{l}{ \ \ \ and from VALD for Si\ii.} \\
	\end{tabular}
\end{table}

\begin{table*}
	\centering
	\caption{NLTE (N) and LTE (L) abundances, $\eps{}$, from lines of Si\ione\ and Si\ii\ in the sample stars.}
	\label{tab:abund_stars}
	\begin{tabular}{lrrlcrcccccccc} % 
		\hline\hline \noalign{\smallskip}
Transition & \multicolumn{1}{c}{ $\lambda$} & log $gf$ & Ref & $\log \Gamma_4/N_{\rm e}$ & & HD    & Vega & HD    & HD     & Sirius & 21 Peg & $\pi$ Cet & $\iota$ Her \\ 
 (\Eexc)          & \multicolumn{1}{c}{ [\AA]}     &          &     &  & & {73666} &      & {72660} & {145788} &        &        &           & \\
\noalign{\smallskip}\hline \noalign{\smallskip}
%(1) & (2) & (3) & (4) & & (5) & (6) &(7) & (8) &(9) & (10) & (11) & (12) \\
%\noalign{\smallskip}\hline \noalign{\smallskip}
\multicolumn{1}{l}{Si\ione} &  & & & & & & &  & & & & & \\
 \eu{3p}{1}{S}{}{0} - \eu{4s}{1}{P}{\circ}{1} & 3905.52 &  $-1$.04 & O91 & $-5.60^1$ & N & 7.54 &       &  7.79 &       &  7.59 &  7.49 &   &       \\
 (1.91~eV)                              &         &        &     & &  L & 7.16 &       &  7.42 &     &  7.26 &  7.08 &   &       \\
\multicolumn{1}{l}{Si\ii} & & & & & & & & & & & &  & \\
 \eu{3p^2}{2}{D}{}{} - \eu{4p}{2}{P}{\circ}{} & 3853.67 &  $-1$.34 & M01 & $-5.20^1$ & N &  7.63 &       &       &      &  7.67 & 7.50 & 7.81 &  7.85   \\
 (6.86~eV)                             &          &       &   &  &  L &  7.64 &       &       &     &  7.66 & 7.50 & 7.49 &  7.25   \\
                                              & 3856.02 &   $-0$.41 & M01 & $-5.24^1$ & N &  7.89 &       &       &     &  7.84 & 7.60 & 7.84 &  8.03   \\
                                          &         &          &   &  &  L &  7.89 &       &       &       &  7.84 & 7.62 & 7.68 &  7.48   \\
                                           & 3862.59 &    $-0$.76 & M01 & $-5.30^1$ & N &  7.94 &       &  7.94 &       &  7.85 & 7.69 & 7.93 &  7.98   \\
                                          &         &          &  &   &  L &  7.95 &       &  7.96 &       &  7.85 & 7.69 & 7.73 &  7.37   \\
 \eu{3d}{2}{D}{}{} - \eu{5p}{2}{P}{\circ}{} & 4075.45 &  $-1$.40 & M01 & $-4.89^2$ & N &  7.52 &       &  7.78 &        &       & 7.53 & 7.65 &  7.67   \\
 (9.84~eV)                           &         &        &  &   &  L &  7.52 &       &  7.77 &       &       & 7.54 & 7.48 &  7.33   \\
                                            & 4076.78 &   $-1$.68 & M01 & $-4.89^2$ & N &       &       &       &       &       &      & 7.71 &         \\
                                           &       &           &  &   &  L &       &       &       &       &       &      & 7.57 &         \\
 \eu{3d}{2}{D}{}{} - \eu{4f}{2}{F}{\circ}{} & 4128.05 &   0.36 & M01 & $-4.96^1$ & N &  7.53 &  6.87 &       &        &  7.64 &      & 7.48 &  7.25   \\
  (9.84~eV)                          &          &       &   &  &  L &  7.59 &  6.95 &       &       &  7.71 &      & 7.47 &  6.84   \\
                                            & 4130.89 &    0.55 & M01 & $-4.91^1$ & N &  7.54 &  6.85 &  7.72 &       &  7.59 & 7.32 & 7.48 &  7.15   \\
                                           & 4130.87 &   $-0$.78 &  &   &  L &  7.61 &  6.93 &  7.78 &       &  7.68 & 7.48 & 7.43 &  6.76   \\
 \eu{4d}{2}{D}{}{} - \eu{7f}{2}{F}{\circ}{} & 4621.42 & $-0$.61 & M95 & $-3.86^3$ & N &       &       &  7.78 &       &       & 7.50 & 7.75 &    b$^4$     \\
(12.52~eV) &        &       &   &  &  L &       &       &  7.76 &       &       & 7.44 &  7.56    &         \\
                                         & 4621.72 &  $-0$.45 & M95 &  & N &       &       &  7.78 &       &       & 7.51 &   7.75   &  7.60    \\
                                        & 4621.70 &  $-1$.75 & M95 & & L &       &       &  7.76 &       &       & 7.46 &  7.56    &  6.95     \\
 \eu{4p}{2}{P}{\circ}{} - \eu{4d}{2}{D}{}{} & 5041.02 &  0.03 & M01 & $-4.80^1$ & N &  7.85 &       &  8.00 &  7.81 &  7.88 & 7.68 & 7.97   &  7.94   \\
  (10.07~eV)                        &       &       &   &    &  L &  7.93 &       &  8.06 &  7.94 &  7.93 & 7.80 & 7.78  &  6.97   \\
                                              & 5055.98 &  0.52 & M01 & $-4.76^1$ & N &  7.65 &  6.92 &  7.77 &  7.60 &  7.75 & 7.45 & 7.79 &  7.66   \\
                                             &         &       &  &   &  L &  7.76 &  6.96 &  7.86 &  7.76 &  7.81 & 7.60 & 7.58 &  6.68   \\
                                              & 5056.32 & $-0$.49 & M01 & $-4.76^1$ & N &  7.73 &       &  7.85 &  7.69 &       & 7.56 &      &  7.78   \\
                                              &         &       &   &  & L &  7.76 &       &  7.88 &  7.74 &       & 7.60 &      &  6.91   \\
 \eu{4d}{2}{D}{}{} - \eu{6f}{2}{F}{\circ}{} & 5466.89 & $-0$.08 & M95 & $-4.20^3$ & N &       &       &       &       &       &        &      &  7.75   \\
 (12.52~eV)                          & 5466.85 & $-1$.38 & M95 & &  L &       &       &       &       &       &       &      &  6.57   \\
 \eu{5p}{2}{P}{\circ}{} - \eu{7d}{2}{D}{}{} & 5469.45 & $-0$.76 & M95 & $-4.06^3$ & N &       &       &       &       &       &        & 7.88 &        \\
 (12.88~eV)                          & 5469.47 & $-1$.72 & M95 & & L &       &       &       &       &       &       & 7.40 &        \\
 \eu{4p}{2}{P}{\circ}{} - \eu{5s}{2}{S}{}{} & 5957.56 & $-0$.22 & B95 & $-4.83^1$ & N &  7.62 &  6.96 &  7.71 &  7.63 &  7.68 & 7.36   & 7.74 &  7.60   \\
 (10.07~eV)                         &       &       &   &  &  L &  7.70 &  7.01 &  7.77 &  7.76 &  7.74 & 7.46    & 7.63 &  6.77   \\
                                              & 5978.93 &  0.08 & B95 & $-4.85^1$ & N &  7.55 &  6.87 &  7.69 &  7.57 &  7.64 & 7.32 & 7.63 &  7.57   \\
                                             &         &       &   &  &  L &  7.65 &  6.93 &  7.78 &  7.74 &  7.72 & 7.48 & 7.57 &  6.74   \\
 \eu{4f}{2}{F}{\circ}{} - \eu{6g}{2}{G}{}{} & 6239.61 &  0.18 & M95 & $-3.54^3$ & N &       &       &  7.85 &       &       & 7.57 & 7.73   &  7.92   \\
 (12.84~eV)                          & 6239.61 & $-1$.12 & M95 & & L &       &       &  7.82 &       &       & 7.41 & 7.33  &  e$^5$  \\
                                           & 6239.66 &  0.02 & M95 &  &    &       &       &       &       &       &      &      &        \\
 \eu{4s}{2}{S}{}{} - \eu{4p}{2}{P}{\circ}{} & 6347.11 &   0.15 & M01 & $-5.08^1$ & N &  7.60 &  6.80 &  7.79 &  7.55 &  7.69 & 7.28 & 7.77  &  8.38    \\
 (8.12~eV)                          &         &       &   &  &  L &  8.20 &  7.32 &  8.31 &  8.14 &  8.14 & 8.06 & 7.99  &  7.78    \\
                                              & 6371.37 & $-0$.08 & M01 & $-5.08^1$ & N &  7.52 &  6.77 &  7.75 &  7.47 &  7.59 & 7.24 & 7.66 &  8.27    \\
                                             &         &       &   &  &  L &  8.09 &  7.12 &  8.15 &  8.04 &  7.95 & 7.89 & 7.84 &  7.60    \\
 \eu{5p}{2}{P}{\circ}{} - \eu{6d}{2}{D}{}{} & 6818.41 & $-0$.52 & M95 & $-4.24^3$ & N &       &       &       &       &       &      &        &  7.94    \\   
(12.88~eV) & 6829.80 & $-0$.26 & M95 &  & N &       &       &       &       &       &      & 7.78 &  7.87    \\
                                             & 6829.83 & $-1$.22 & M95 & & L &       &       &       &       &       &      & 7.34 &   e     \\
 \eu{4d}{2}{D}{}{} - \eu{5f}{2}{F}{\circ}{} & 7848.82 &  0.32 & M95 & $-4.68^3$ & N &       &       &       &       &       & 7.52 & 7.55   &  8.36   \\
 (12.52~eV)                          &         &       &   &  &  L &       &       &       &       &       & 7.32 & 7.14  &  e      \\
                                              & 7849.72 &  0.47 & M95 & & N &       &       &       &       &       & 7.48 & 7.64 &  8.29   \\
                                             & 7849.62 & $-0$.83 & M95 & & L &       &       &       &       &       & 7.28 & 7.23 &  e       \\
\noalign{\smallskip}\hline \noalign{\smallskip}
\multicolumn{14}{l}{{\bf Notes.} $\Gamma_4/N_{\rm e}$ in rad/s$\cdot$cm$^3$, $^1$ \citet{2009A&A...508..491B}; $^2$ Wilke (2003, Ph.D. Thesis) as given by \citet{2009AA...503..945F};} \\
\multicolumn{14}{l}{$^3$  VALD; $^4$ blend; $^5$ emission line.} \\
\multicolumn{14}{l}{{\bf Ref.} B95 = \citet{1995PhyS...52..628B}, M01 = \citet{2001JQSRT..69..535M}, M95 = \citet{1995JPhB...28.3485M}, O91 = \citet{1991PhRvA..44.7134O}.} \\
%\multicolumn{14}{l}{O91 = \citet{1991PhRvA..44.7134O}.}
\noalign{\smallskip}\hline
	\end{tabular}     
\end{table*}              

%, Heinrich-Heine-Universit{\"a}t, D{\"u}sseldorf)
\begin{table}
	\centering
	\caption{NLTE and LTE abundances, $\eps{}$, from lines of Si\iii\ in $\pi$~Cet and $\iota$~Her.}
	\label{tab:abund_si3}
	\begin{tabular}{ccrccccc} % 
		\hline\hline \noalign{\smallskip}
\multicolumn{1}{c}{ $\lambda$} & \Eexc & log $gf$ & \multicolumn{2}{c}{$\pi$ Cet} & \multicolumn{2}{c}{$\iota$ Her} \\ 
\cline{4-5}
\cline{7-8}
\multicolumn{1}{c}{ [\AA]}     & [eV]  &          & NLTE  & LTE & & NLTE  & LTE \\
		\hline\hline \noalign{\smallskip}
4552.62 & 19.02 &  0.29 &      &      & & 7.47 & 7.82 \\
4567.84 & 19.02 &  0.07 & 7.69 & 7.82 & & 7.52 & 7.80 \\
4574.76 & 19.02 & $-0$.41 & 7.74 & 7.82 & & 7.52 & 7.70 \\
5739.73 & 19.72 & $-0$.10 &      &      & & 7.63 & 7.81 \\
\noalign{\smallskip}\hline \noalign{\smallskip}
\multicolumn{8}{l}{{\bf Note.} $gf$-values are from NIST.}
	\end{tabular}     
\end{table}              

\subsection{Non-LTE effects on spectral lines}
%\subsection{Stellar silicon abundances}

\begin{figure}
	% To include a figure from a file named example.*
	% Allowable file formats are eps or ps if compiling using latex
	% or pdf, png, jpg if compiling using pdflatex
	\includegraphics[width=\columnwidth]{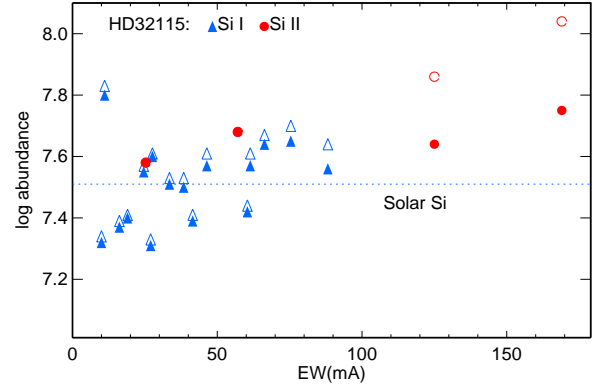}
    \caption{NLTE (filled symbols) and LTE (open symbols) abundances from lines of Si\ione\ (triangles) and Si\ii\ (circles) in HD~32115. }
    \label{fig:hd32115}
\end{figure}

\begin{figure*}
 \begin{minipage}{150mm}

\hspace{-10mm}
\parbox{0.45\linewidth}{\includegraphics[scale=0.55]{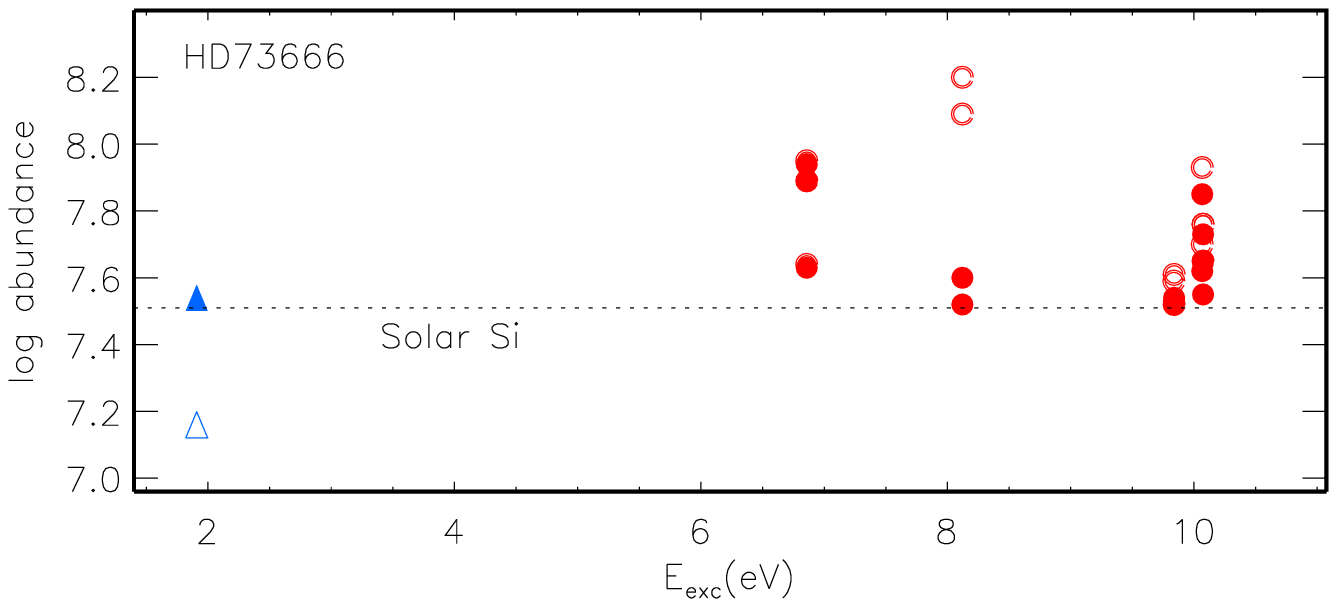}\\
\centering}
%\hspace{0.2\linewidth}
\hspace{10mm}
\parbox{0.45\linewidth}{\includegraphics[scale=0.55]{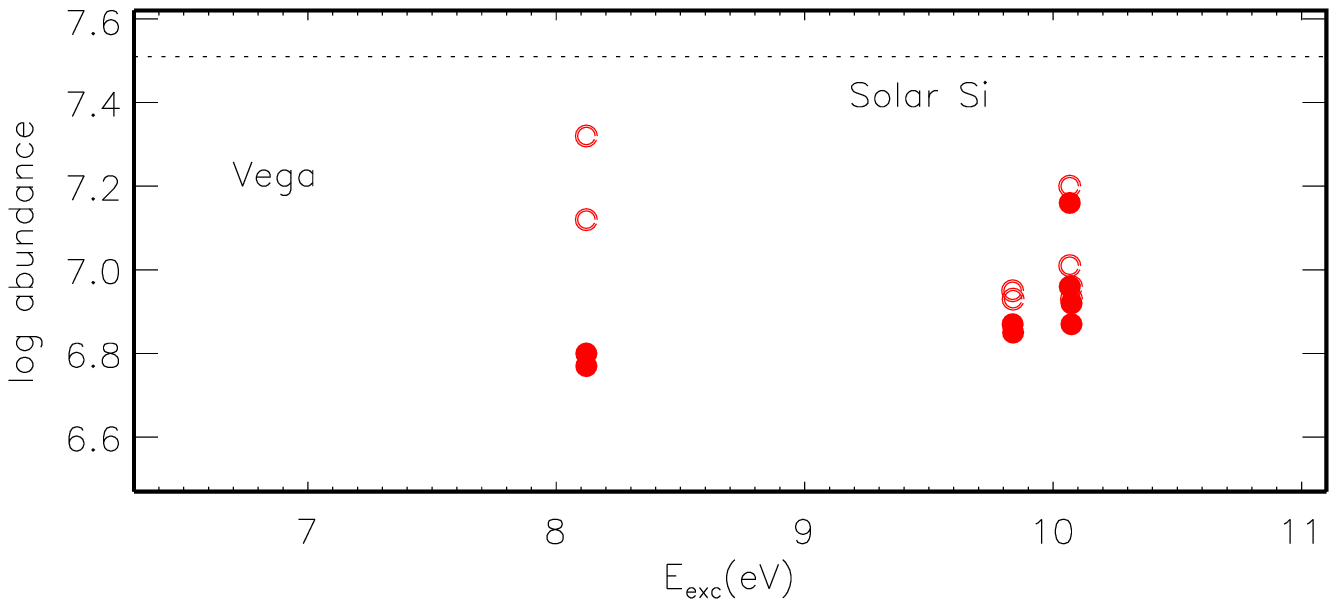}\\
\centering}
\hfill
\\[0ex]

\vspace{-7mm}
\hspace{-10mm}
\parbox{0.45\linewidth}{\includegraphics[scale=0.55]{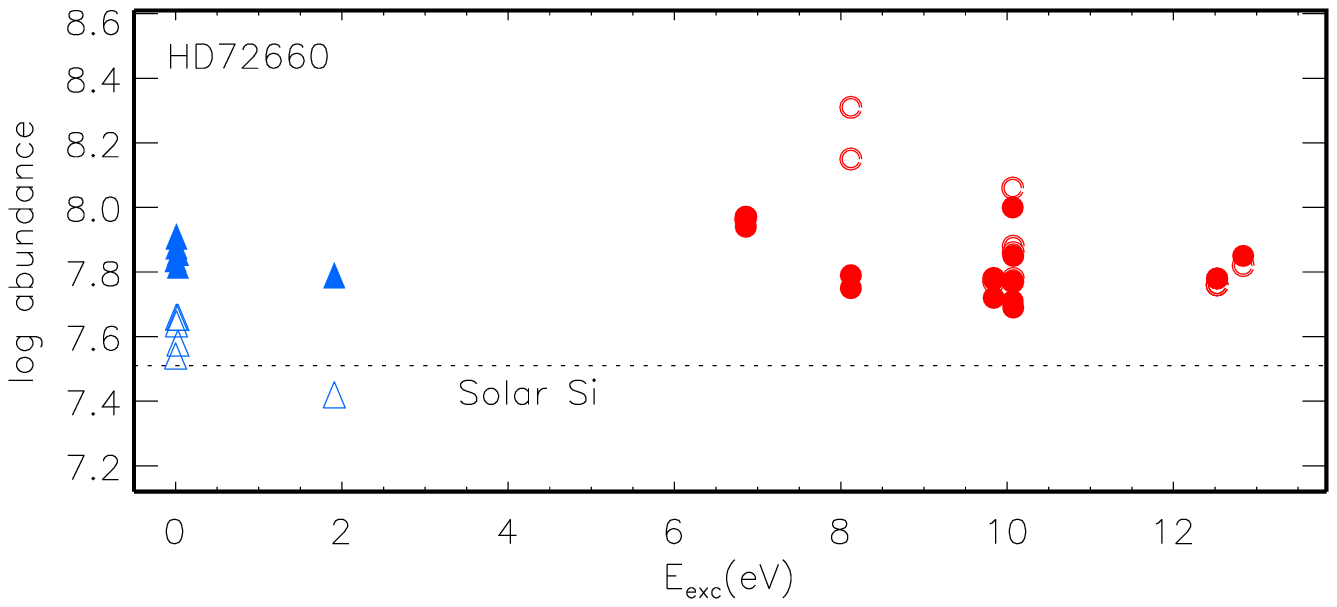}\\
\centering}
%\hspace{1\linewidth}
\hspace{10mm}
\parbox{0.45\linewidth}{\includegraphics[scale=0.55]{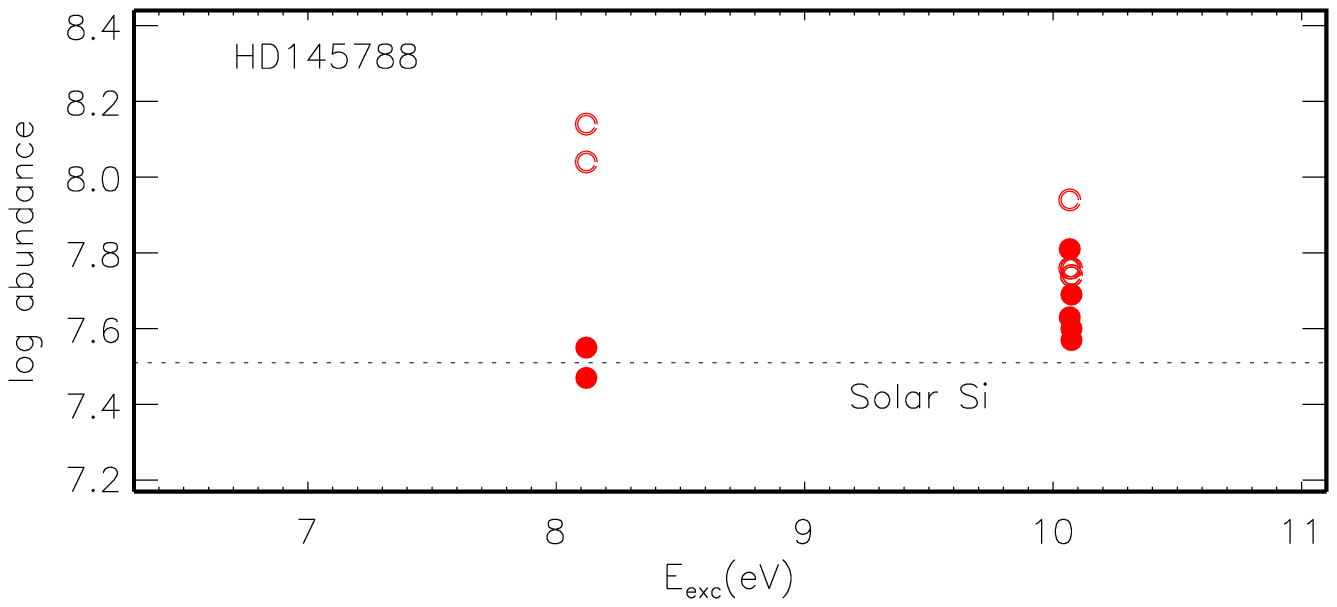}\\
\centering}
\hfill
\\[0ex]

\vspace{-7mm}
\hspace{-10mm}
\parbox{0.45\linewidth}{\includegraphics[scale=0.55]{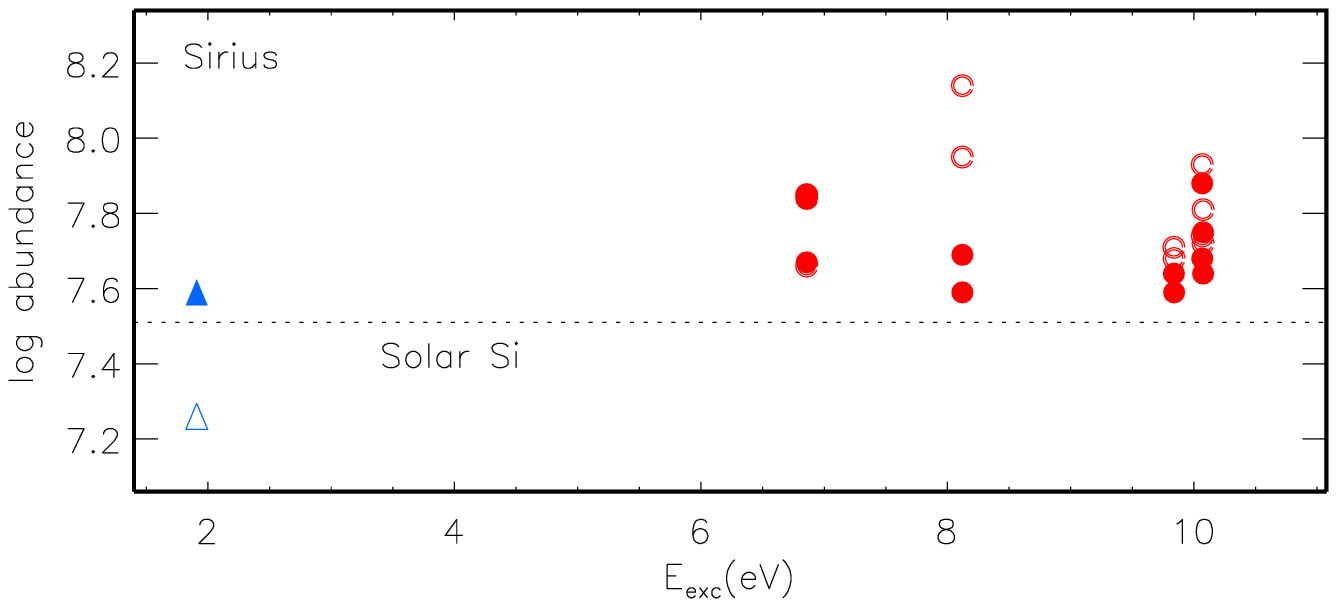}\\
\centering}
%\hspace{1\linewidth}
\hspace{10mm}
\parbox{0.45\linewidth}{\includegraphics[scale=0.55]{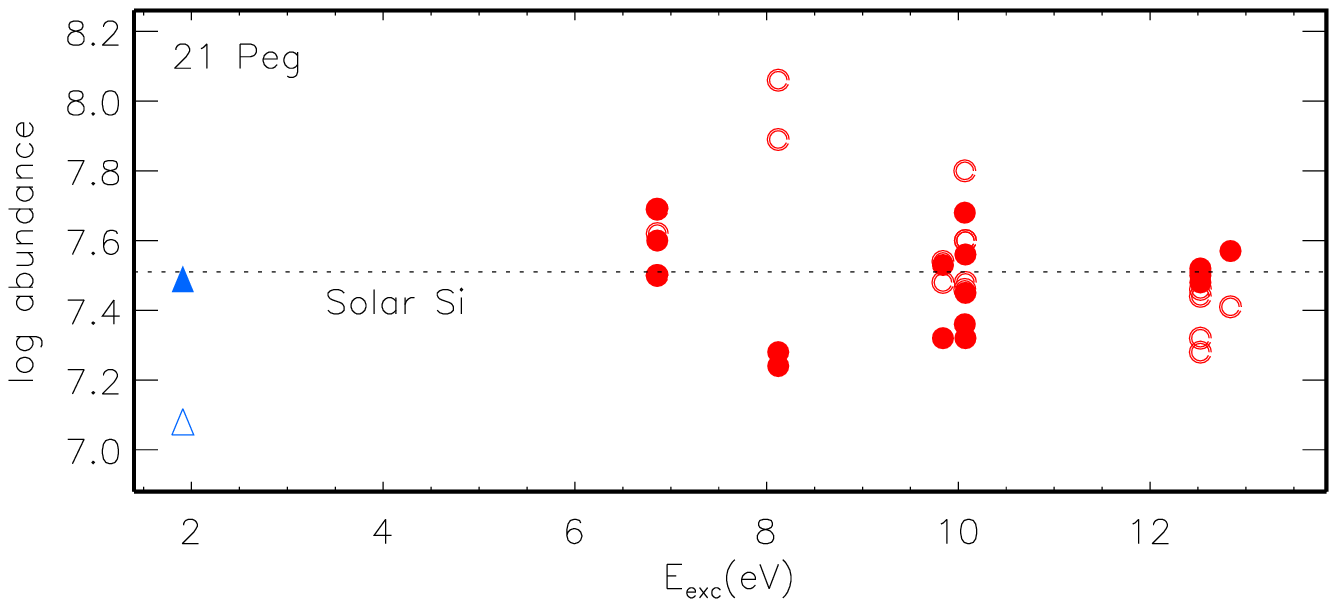}\\
\centering}
\hfill
\\[0ex]

\vspace{-7mm}
\hspace{-10mm}
\parbox{0.45\linewidth}{\includegraphics[scale=0.55]{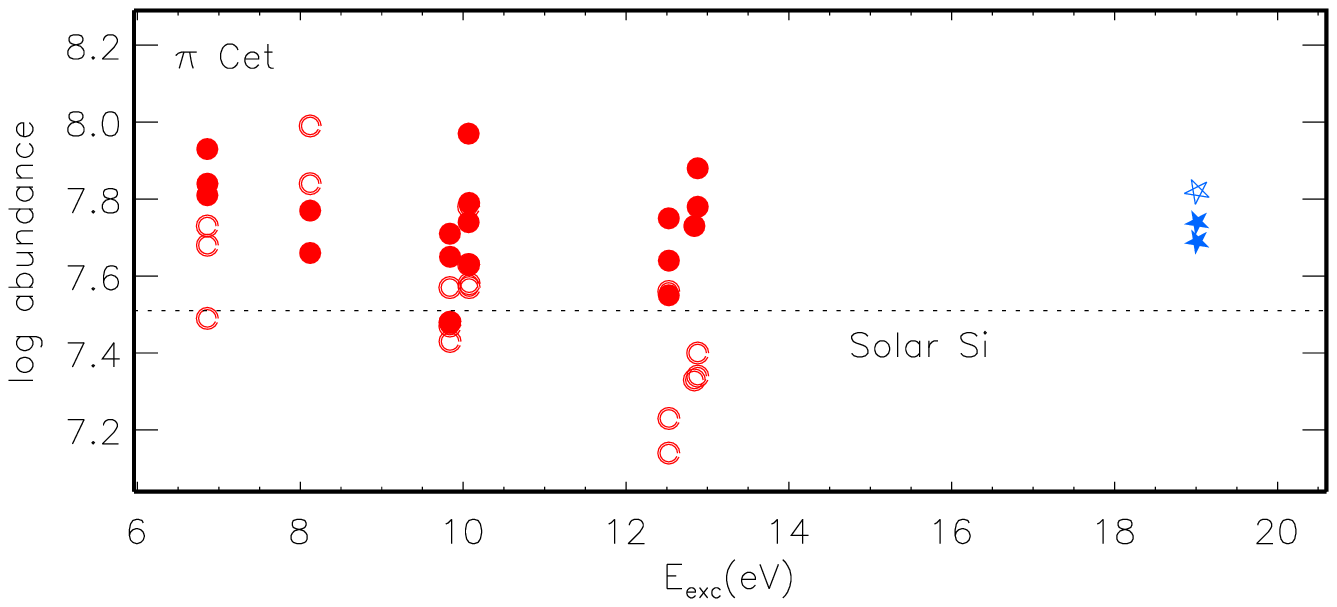}\\
\centering}
%\hspace{1\linewidth}
\hspace{10mm}
%\parbox{0.45\linewidth}{\includegraphics[scale=0.55]{si23_iHer.ps}\\
%\parbox{0.45\linewidth}{\includegraphics[scale=0.55]{si23nohydnlte_exc_iHer.ps}\\
\parbox{0.45\linewidth}{\includegraphics[scale=0.55]{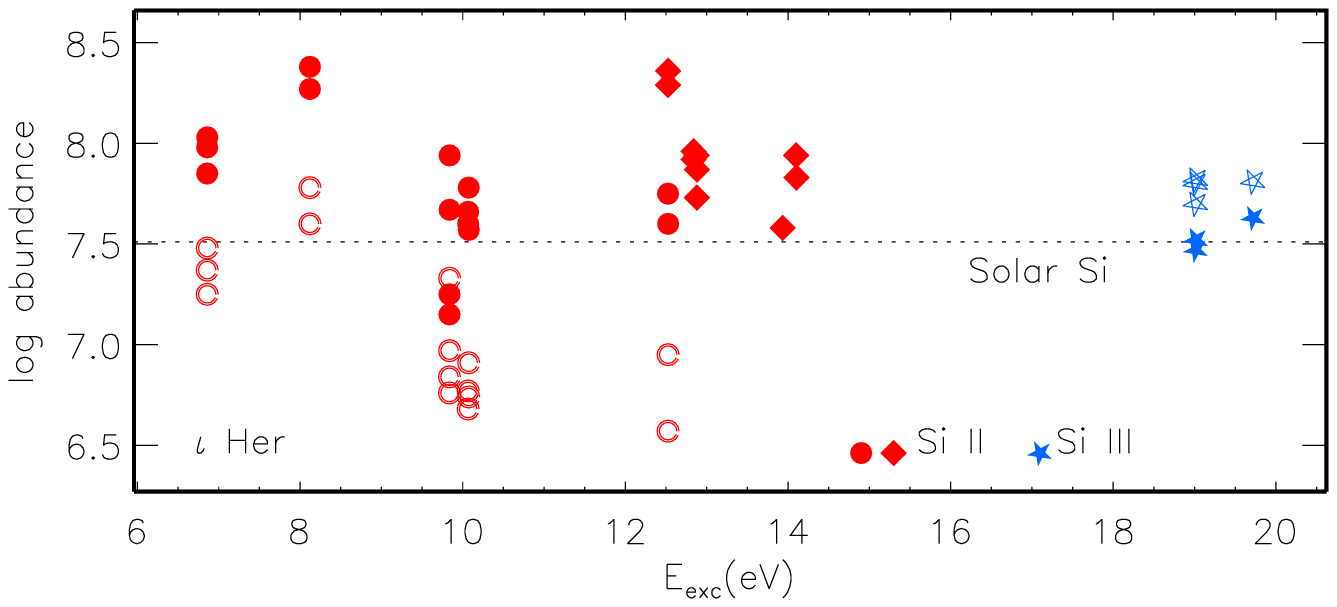}\\
\centering}
\hfill
\\[0ex]
	% To include a figure from a file named example.*
	% Allowable file formats are eps or ps if compiling using latex
	% or pdf, png, jpg if compiling using pdflatex
%	\includegraphics[width=\columnwidth]{si12_exc_hd73666_7.ps}
%	\includegraphics[width=\columnwidth]{si2_exc_vega_7.ps}
%	\includegraphics[width=\columnwidth]{si12_exc_hd72660_7.ps}
%	\includegraphics[width=\columnwidth]{si2_exc_hd145788_7.ps}
%	\includegraphics[width=\columnwidth]{si12_exc_sirius_7.ps}
%	\includegraphics[width=\columnwidth]{si12_exc_piCet.ps}
%	\includegraphics[width=\columnwidth]{si23_exc_iHer.ps}
    \caption{NLTE (filled symbols) and LTE (open symbols) abundances from individual lines of Si\ione\ (triangles), Si\ii\ (circles), and Si\iii\ (5 pointed stars) in the sample stars. For $\iota$~Her, the NLTE abundances derived from the Si\ii\ emission lines are shown by the rhombi. The dotted line indicates the solar system silicon abundance, $\eps{\odot}$ = 7.51$\pm$0.01   \citep{2019arXiv191200844L}.}
    \label{fig:hd72660_picet}
\end{minipage}
\end{figure*}

%Both LTE and NLTE abundances were determined from line profile fitting.
Tables~\ref{tab:sun_hd32115}, \ref{tab:hd72660}, \ref{tab:abund_stars}, and \ref{tab:abund_si3} and Figs.~\ref{fig:hd32115} and \ref{fig:hd72660_picet} present the NLTE and LTE abundances from individual lines in the sample stars. 

\subsubsection{Lines of Si\ione}

The 16 lines of Si\ione\ were measured in our coolest star, HD~32115 (Table~\ref{tab:sun_hd32115}, Fig.~\ref{fig:hd32115}). They all arise from the \Eexc\ $<$ 6.2~eV levels, for which b $>$ 1 in the line-formation layers, while populations of their upper levels are lower than the LTE ones. Therefore, the lines are strengthened in NLTE, and the NLTE abundance corrections, $\Delta_{\rm NLTE} = \eps{NLTE} - \eps{LTE}$, are negative. However, the NLTE effects are small, so that $\Delta_{\rm NLTE}$ ranges between $-0.02$ and $-0.06$~dex for different lines. 

\begin{figure}
	% To include a figure from a file named example.*
	% Allowable file formats are eps or ps if compiling using latex
	% or pdf, png, jpg if compiling using pdflatex
	\includegraphics[width=\columnwidth]{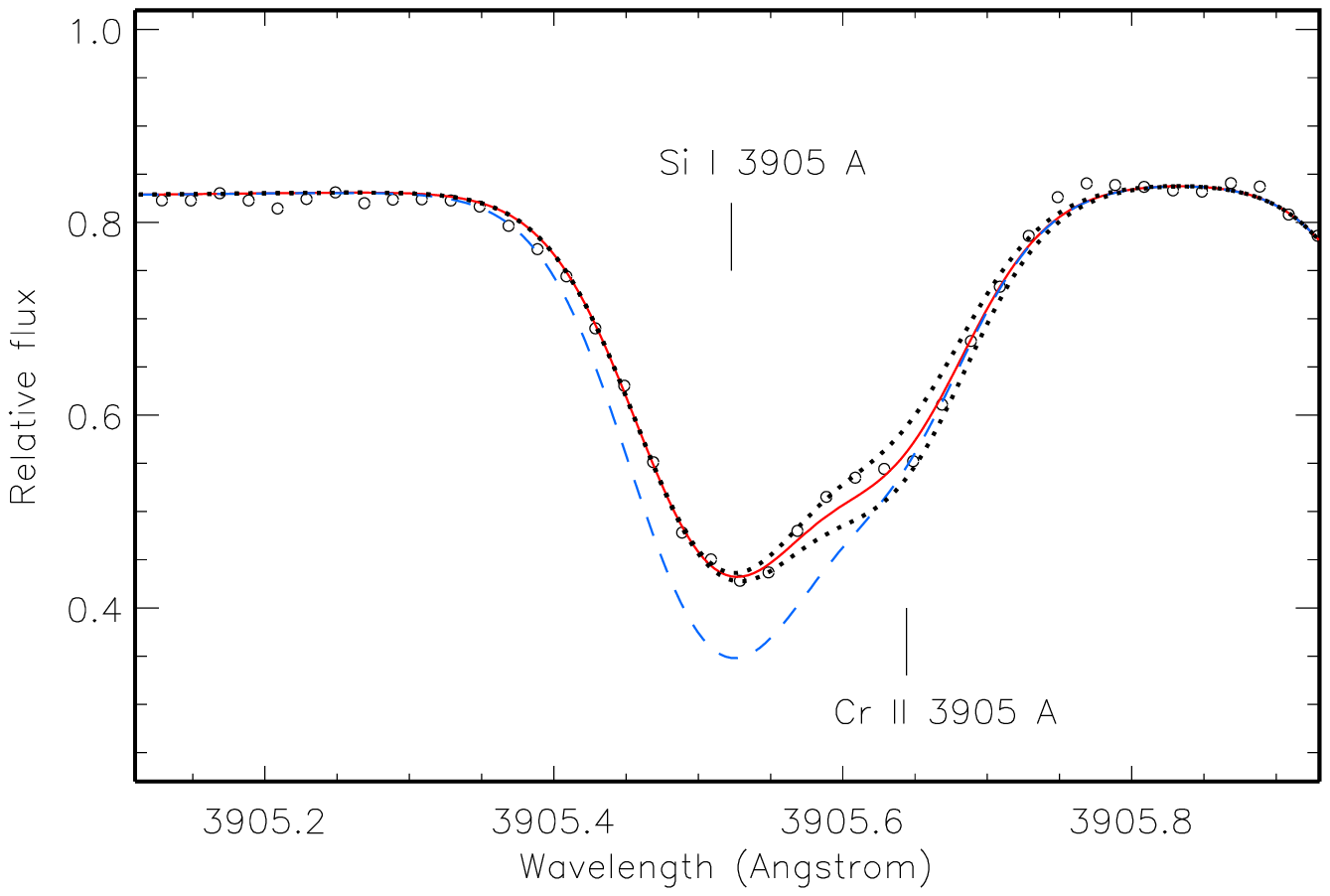}
    \caption{Best NLTE fit (solid curve) of Si\ione\ 3905\,\AA\ in HD~72660 (open circles).  The corresponding LTE line profile is plotted by the dashed curve. The dotted curves show the NLTE synthetic spectra computed with 0.1~dex higher and lower abundance of Cr. }
    \label{fig:si3905}
\end{figure}

In the hotter, 9380 $\le \Teff \le$ 10400~K, stars, only Si\ione\ 3905\,\AA\ was measured in the visible spectral range. We note that observed spectra of Vega and HD~145788 do not cover this line. Overionisation of Si\ione\ leads to weakened line (Fig.~\ref{fig:si3905}) and large positive NLTE abundance correction, which ranges between 0.33 and 0.41~dex for different stars. Si\ione\ 3905.523\,\AA\ is blending with the Cr\ii\ 3905.644\,\AA\ line. However, as shown in Fig.~\ref{fig:si3905}, a variation of 0.1~dex in the Cr abundance does not influence the derived silicon abundance. 

We are lucky to measure five lines of Si\ione\ in the UV spectrum of HD~72660 (Table~\ref{tab:hd72660}). Their NLTE abundances agree well with that from Si\ione\ 3905\,\AA, so that the dispersion in 
the single line measurements around the mean, $\sigma = \sqrt{\Sigma(\overline{x}-x_i)^2 / (N_l-1)}$, amounts to 0.04~dex (Table~\ref{tab:stars_param}). Here, $N_l$ is the number of measured lines. This provides evidence for a reliable abundance from Si\ione\ 3905\,\AA\ despite blending with Cr\ii\ 3905\,\AA. 
%The greater line-to-line scatter for Si\ione\ is observed in LTE, with $\sigma$ = 0.09~dex.

\subsubsection{Absorption lines of Si\ii}

The Si\ii\ multiplets in Table~\ref{tab:abund_stars} can be separated into three groups with respect to the departures from LTE.

1. Si\ii\ 3853-62\,\AA, 4075-76\,\AA, 4128-30\,\AA, 5041-56\,\AA, and 5957-78\,\AA. In the model atmospheres with $\Teff \le$ 10400~K, where Si\ii\ is a majority species, the lower and upper levels of the corresponding transitions are tightly coupled to the ground state in the line-formation region, below log~$\tau_{5000} \simeq -2$ (see Fig.~\ref{fig:bfactors}). As a result, the NLTE effects are small, $\Delta_{\rm NLTE}$ is slightly negative, and does not exceed 0.02~dex for the two bluest multiplets and 0.16~dex for the remaining ones. In the atmospheres of $\pi$~Cet and $\iota$~Her, all these lines are weakened due to overionisation of Si\ii\ and the NLTE abundance corrections are positive. For example, they amount to 0.34 to 0.98~dex for different lines in $\iota$~Her.

\begin{figure}
	% To include a figure from a file named example.*
	% Allowable file formats are eps or ps if compiling using latex
	% or pdf, png, jpg if compiling using pdflatex
	\includegraphics[width=\columnwidth]{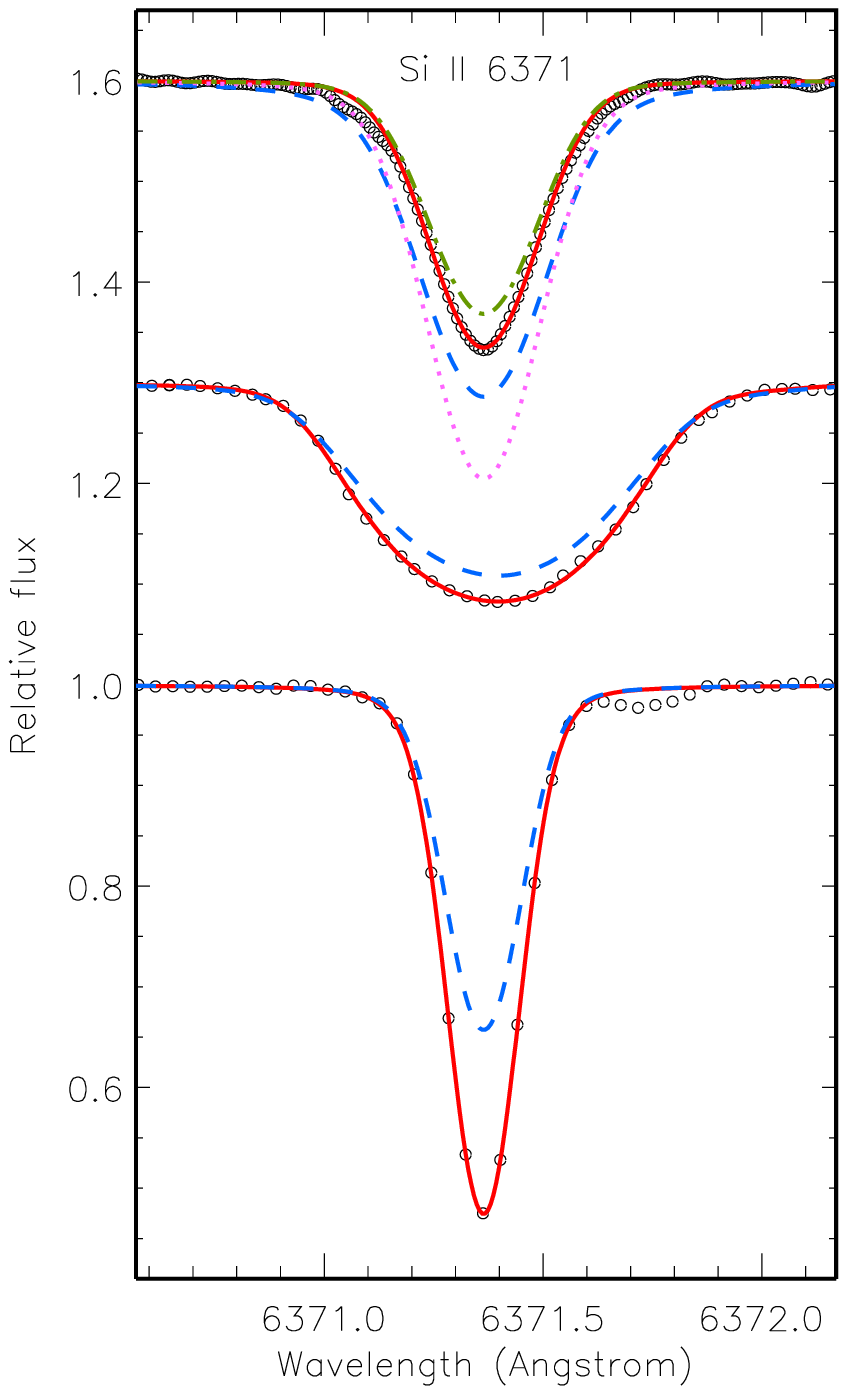}
    \caption{Best NLTE fits (solid curve) of Si\ii\ 6371\,\AA\ in $\iota$~Her, $\pi$~Cet, and 21~Peg (open circles, from top to down). The corresponding LTE line profiles are plotted by the dashed curves. For $\iota$~Her, we also show the NLTE line profiles computed by replacing electron-impact excitation data of \citet{2014MNRAS.442..388A} with the \citet{Reg1962} approximation for the allowed transitions and $\Upsilon$ = 1 for the forbidden transitions (dash-dotted curve) and by replacing the TOPbase photoionisation cross sections for the Si\ii\ \eu{4s}{2}{S}{}{}, \eu{3d}{2}{D}{}{}, and \eu{3p^2}{2}{D}{}{} levels with the hydrogenic ones (dotted curve). For better visibility, spectra of $\pi$~Cet and $\iota$~Her are shifted along Y axis.}
    \label{fig:si6371}
\end{figure}

2. Si\ii\ 6347, 6371\,\AA. In the model atmospheres with $\Teff \le$ 12800~K, NLTE leads to strengthened lines (Fig.~\ref{fig:si6371} for 6371\,\AA) owing to dropping the line source function ($S_\nu$) below the Planck function ($B_\nu$) in the line-formation layers (see Fig.~\ref{fig:bfactors}). The NLTE effects grow toward higher effective temperature, so that, for Si\ii\ 6371\,\AA, a magnitude of $\Delta_{\rm NLTE}$ increases from $-0.22$~dex in HD~32115 to $-0.65$~dex in 21~Peg. In the atmosphere of $\pi$~Cet, overionisation of Si\ii\ competes with dropping $S_\nu / B_\nu$, however, the latter effect prevails, resulting in $\Delta_{\rm NLTE} = -0.18$~dex. Overionisation of Si\ii\ is the dominant NLTE mechanism in the atmosphere of the hottest star, so that the Si\ii\ 6347, 6371\,\AA\ lines are greatly weakened and $\Delta_{\rm NLTE} = 0.60$ and 0.67~dex, respectively.

\begin{figure}
	% To include a figure from a file named example.*
	% Allowable file formats are eps or ps if compiling using latex
	% or pdf, png, jpg if compiling using pdflatex
	\includegraphics[width=\columnwidth]{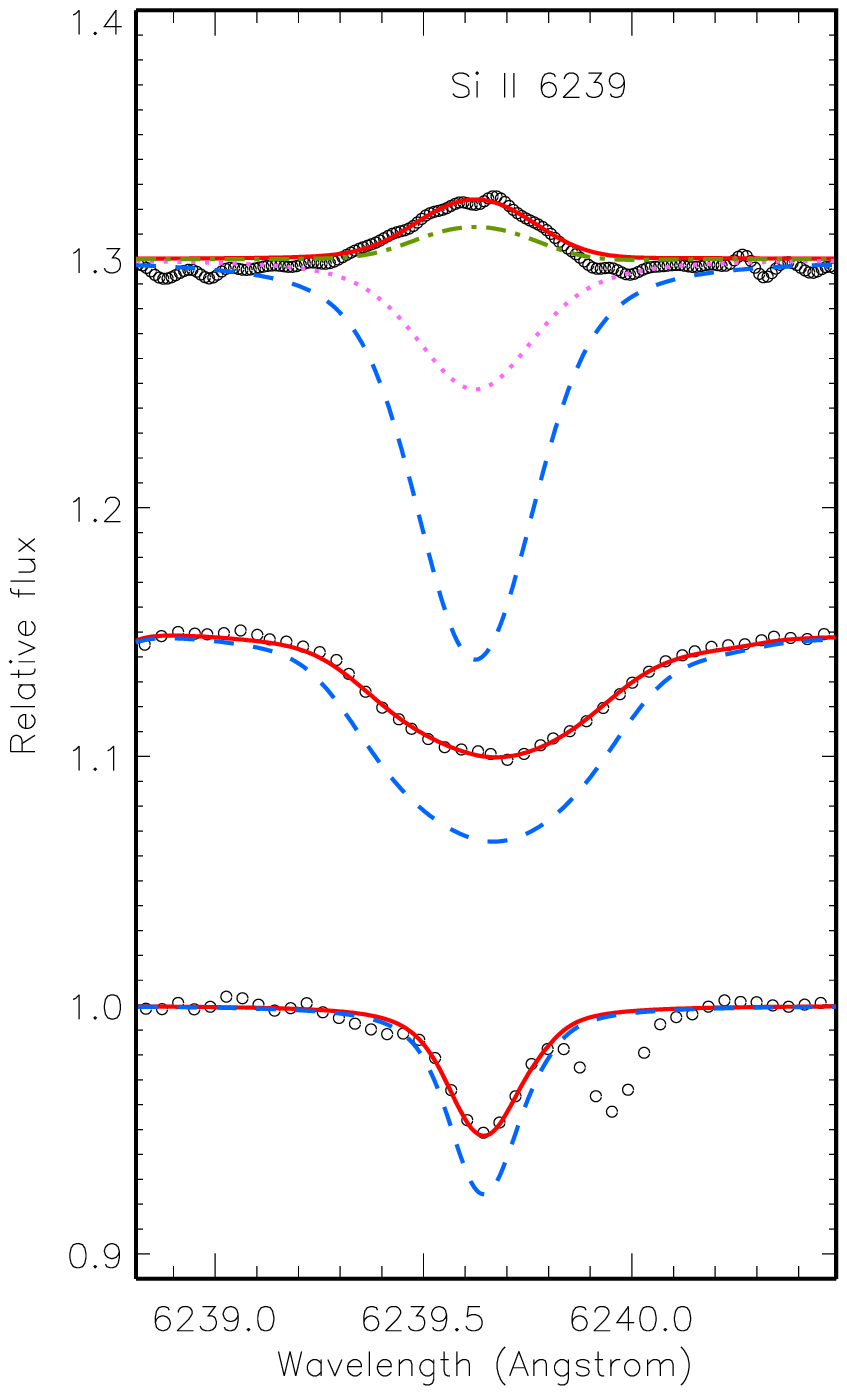}
    \caption{The same as in Fig.\,\ref{fig:si6371} for Si\ii\ 6239\,\AA. }
    \label{fig:si6239}
\end{figure}

3. The high-excitation (\Eexc\ $>$ 12~eV) lines are observed in the three hottest stars. The exception is Si\ii\ 4621.4, 4621.7\,\AA\ and 6239\,\AA\ in HD~72660, which were measured thanks to the narrower lines compared with that in the other stars of close effective temperature. The high-excitation levels are depopulated in the line-formation layers due to photon loss in the transitions to the low-lying levels in the $\Teff \le$ 10400~K model atmospheres and due to overionisation of Si\ii\ in the hotter atmospheres. For each transition, the lower level is depopulated to a greater extent than is the upper level, resulting in weakened line compared with its LTE strength (Fig.~\ref{fig:si6239} for 6239\,\AA). The NLTE effects grow toward higher $\Teff$. For example, for Si\ii\ 6239\,\AA, $\Delta_{\rm NLTE}$ = 0.03, 0.16, and 0.40~dex for HD~72660, 21~Peg, and $\pi$~Cet, respectively, and the line appears in emission in $\iota$~Her. In this hottest star of our sample, only 4621.7\,\AA\ and 5466.8\,\AA\ among the high-excitation lines are observed in absorption, although they are greatly weakened by the NLTE effects, with $\Delta_{\rm NLTE}$ = 0.65 and 1.18~dex, respectively. The lines on the long-wave side of 6239\,\AA\ either come into emission or disappear. They deserve a special consideration in Sect.~\ref{sect:emission}.

\subsubsection{Lines of Si\iii}

Observed lines of Si\iii\ (Table~\ref{tab:abund_si3}) arise from the transitions \eu{4s}{3}{S}{}{} -- \eu{4p}{3}{P}{\circ}{} (\Eexc\ = 19.02~eV) and \eu{4s}{1}{S}{}{} -- \eu{4p}{1}{P}{\circ}{} (\Eexc\ = 19.72~eV). In $\pi$~Cet, only the first multiplet was measured, the lines are weak, and form in deep atmospheric layers, around log~$\tau_{5000} \sim -0.1$ (Fig.~\ref{fig:bfactors}), where the NLTE effects on the lines are caused by slight overpopulation of the lower level relative to its TE population. For Si\iii\ 4567 and 4574\,\AA, $\Delta_{\rm NLTE}$ = $-0.13$ and $-0.08$~dex, respectively.

In $\iota$~Her, the Si\iii\ lines are stronger and form in the layers, where the lower levels of both multiplets have enhanced excitation owing to pumping UV transitions from the low-excitation levels, while the upper levels are depopulated via spontaneous transitions to the low-lying levels. Increasing the line absorption coefficient due to b$_{low} > 1$ and dropping the line source function below the Planck function due to b$_{up}$/b$_{low} < 1$ result in strengthened lines and negative $\Delta_{\rm NLTE}$ of $-0.18$ to $-0.35$~dex for different lines.

\subsection{Comparison with other NLTE studies}

The obtained NLTE abundances can only be compared with the results of \citet{2001A&A...373..998W} for lines of Si\ii\  in Vega and \citet{2012A&A...539A.143N} for lines of Si\iii\ in $\iota$~Her. For the three Si\ii\ lines in common, namely, at 4128, 4130, and 5055\,\AA, the NLTE abundance corrections computed by \citet{2001A&A...373..998W} do not exceed 0.11~dex, in absolute value, and agree within 0.03~dex with ours. The NLTE abundances obtained in this study from individual lines of Si\iii\ in $\iota$~Her agree within 0.03 to 0.08~dex with those of \citet{2012A&A...539A.143N}, and the mean abundances are consistent within 0.04~dex in the two studies.

%use common gf nist
%Vega: M2019   W2001
%4128  6.87    6.87
%     -0.08   -0.09
%4130  6.85    6.86
%     -0.08   -0.11
%5055  6.92    7.04   
%     -0.04   -0.06
%use Nieva, Przybilla, 2012 for comparison (i Her)  mean = 7.50

%\section{Stellar silicon abundances, Si\ione /Si\ii, and Si\ii /Si\iii\ ionisation equilibrium}
\section{Abundance analyses of the $\Teff \le$ 12800~K stars}\label{sect:mild}

For each of the eight stars with $\Teff \le$ 12800~K, its silicon spectrum is well reproduced in NLTE with a unique element abundance. The star $\iota$~Her, with both absorption and emission lines in its spectrum, is discussed in the next section. The average NLTE and LTE abundances from each ionisation stage observed in a given star are presented in Table~\ref{tab:stars_param}. 

We comment on the Si\ione\ lines in HD~32115. They reveal a substantial scatter of abundances, with $\sigma$ = 0.17~dex, independent of either LTE or NLTE and independent of either measured equivalent widths (Fig.~\ref{fig:hd32115}) or excitation potentials. 
%The most extended sample of Si\ione\ lines was measured in our coolest star, HD~32115 (Table~\ref{tab:sun_hd32115}, ).  We note rather large line-to-line scatter, with $\sigma$ = 0.17~dex, independent of either LTE or NLTE. 
Such a scatter is, most probably, due to the uncertainties in $gf$-values. Indeed, the dispersion was substantially reduced, down to $\sigma$ = 0.06~dex, in a line-by-line differential approach, where from stellar line abundances we subtracted individual abundances of their solar counterparts. The solar abundances (Table~\ref{tab:sun_hd32115}) were derived using the Kitt Peak Solar Flux Atlas \citep{Atlas} and the calculations with the MARCS model atmosphere
5777/4.44/0 \citep{Gustafssonetal:2008} and a depth-independent microturbulence of 0.9\,\kms. 
%In both non-differential, and differential approach, we obtain consistent, within 0.07~dex, NLTE abundances from lines of Si\ione\ and Si\ii\ in HD~32115, while the difference in LTE abundances amounts to $-0.21$~dex.

For Si\ii\ in each star, NLTE reduces substantially the line-to-line scatter compared with the LTE case. For example, from $\sigma$ = 0.24 to 0.08~dex for HD~32115 (four Si\ii\ lines) and from $\sigma$ = 0.25 to 0.14~dex for $\pi$~Cet (19 lines). 

The NLTE line formation is essential for achieving consistent abundances from Si\ione\ and Si\ii\ in the five stars, where lines of both ionisation stages were measured. For example, the abundance difference (Si\ione\ -- Si\ii) = $-0.07$ and $-0.21$~dex in NLTE and LTE, respectively, for HD~32115, and the corresponding numbers are $-0.01$ and $-0.54$~dex for 21~Peg. Fairly consistent NLTE abundances from lines of Si\ii\ and Si\iii\ were found in $\pi$~Cet, while, in LTE, the abundance difference amounts to $-0.23$~dex. 

The [Si/H] values were computed using the solar Si abundance, $\eps{\odot}$ = 7.51$\pm$0.01, as recommended by \citet{2019arXiv191200844L}. For stellar Si abundance, we computed the average value, if silicon was observed in the two ionisation stages. The exception is Sirius. We relied on its Si\ii\ based abundance, having in mind that the only measured Si\ione\ line is affected by the Cr\ii\ line (see Fig.~\ref{fig:si3905} for HD~72660) and Sirius is an Am star. 
We obtained that the silicon abundance follows the iron one in our sample stars, including a $\lambda$~Boo star Vega and the three Fe-rich stars, 
%HD~72660, HD~145788, and Sirius, 
but except for $\pi$~Cet. This suggests that the mechanisms, which produced deviations in metal abundances of our chemically peculiar stars from the solar one, did not separate chemical elements. 
%four stars have non-solar iron abundances, namely, , with [Fe/H] = $-0.50$, and with [Fe/H] $\sim 0.4$.
%As expected, a $\lambda$~Boo star Vega reveals a deficiency of silicon and our  have supersolar Si abundances. 
%A moderate enhancement in silicon, with [Si/H] = 0.12, was found for our third Fe-rich star. 
For $\pi$~Cet, with [Fe/H] = 0, according to \citet{2009AA...503..945F}, we found a supersolar abundance of [Si/H] = 0.23 from lines of both ionisation stages, Si\ii\ and Si\iii.
% was found with a confidence, from the two ionisation stages, for $\pi$~Cet. The Si abundances of the remaining three stars are consistent within the error bars with the solar value.

\begin{table*}
	\centering
	\caption{Error estimates for the NLTE calculations of the silicon lines in $\iota$~Her.}
	\label{tab:uncertainties}
	\begin{tabular}{llccccc} % 
	\hline\hline \noalign{\smallskip}
 & & \multicolumn{5}{c}{Changes in $\eps{Si}$ (dex)} \\
\cline{3-7}
 & & \multicolumn{4}{c}{Si\ii } & Si\iii \\
 & & 3856\,\AA & 5978\,\AA & 6239\,\AA & 6371\,\AA & 4567\,\AA \\
\noalign{\smallskip}\hline \noalign{\smallskip}
 Atmospheric parameters: & \multicolumn{6}{c}{ } \\
 $\Teff$ -- 200~K & $\sigma_{\Teff}$ & $-0.02$ & $-0.05$ & +0.07 & $-0.06$ & +0.06 \\
 $\logg$ + 0.05    & $\sigma_{\logg}$ & $-0.02$ & $-0.01$ & +0.02 & $-0.01$ & +0.04 \\
 $\xi_t$ + 0.5~\kms & $\sigma_{\xi}$  & $-0.08$ & $-0.01$ & 0.00  & $-0.06$ & $-0.04$ \\
 Line data: $\Gamma_4$ * 0.5 & $\sigma_{\Gamma}$ & +0.05 & +0.01 & 0.00 & +0.04 & +0.05 \\
 Photoionisations:  & \multicolumn{6}{c}{ } \\
 cross sections * 1.1 & $\sigma_{\rm RBF}$ & 0.00 & $-0.01$ & +0.01 & 0.00 & 0.00 \\
 cross sections * 2     &                    & $-0.02$ & $-0.03$ & +0.02 & $-0.01$ & 0.00 \\
 Si\ii\ \eu{3p}{2}{P}{\circ}{} -- \eu{3p^2}{2}{D}{}{}, & \multicolumn{6}{c}{} \\
%from the ground state to &     \multicolumn{6}{c}{ } \\           
 \eu{3p^2}{2}{S}{}{}, \eu{4s}{2}{S}{}{} transitions: & \multicolumn{6}{c}{ } \\
$f_{lu}$ * 10  &        & +0.16 & $-0.04$ & +0.03 & $-0.33$ & 0.00 \\
 Collisional transitions: & \multicolumn{6}{c}{ } \\
 ($\Upsilon$ and $C_{vR}$) * 2 &                & $-0.01$ & +0.03 & abs & 0.00 & 0.00 \\
 ($\Upsilon$ and $C_{vR}$) * 0.5 & $\sigma_{\rm CBB}$ & +0.01 & $-0.05$ & $-0.50$ & 0.00 & 0.00 \\
 ($\Upsilon$, $C_{vR}$, and $C_{AK}$) * 2 &    & $-0.02$ & $-0.03$ & abs & $-0.07$ & 0.00 \\
\noalign{\smallskip}\hline \noalign{\smallskip}
%Total uncertainty        & $\sigma_{\rm sys}$ & 0.09 & 0.07 & 0.51 & 0.09 & 0.08 \\
%\noalign{\smallskip}\hline \noalign{\smallskip}
 $\Delta_{\rm NLTE}$ & & 0.55 & 0.83 & e & 0.67 & $-0.28$ \\
\noalign{\smallskip}\hline \noalign{\smallskip}
\multicolumn{7}{l}{{\bf Notes.} $C_{vR}$ = \citet{Reg1962} rates; $C_{AK}$ = \citet{2014MNRAS.442..388A} rates; } \\
\multicolumn{7}{l}{abs = absorption is predicted; e = emission line; 0.00 means smaller than 0.01, in absolute value. }
	\end{tabular}     
\end{table*}              

\begin{table}
	\centering
	\caption{Emission lines of Si\ii\ in $\iota$~Her.}
	\label{tab:emission_si2}
	\begin{tabular}{lccrc} % 
	\hline\hline \noalign{\smallskip}
 Transition & \Eexc & \multicolumn{1}{c}{ $\lambda$} & log $gf$ & $\eps{}$ \\ 
            & [eV]  & \multicolumn{1}{c}{ [\AA]}     &          &  NLTE  \\
\noalign{\smallskip}\hline \noalign{\smallskip}
 \eu{4f}{2}{F}{\circ}{} - \eu{6g}{2}{G}{}{} & 12.84 & 6239.61 &  0.18 &  7.92  \\
                                            &       & 6239.61 & -1.12 &        \\
                                            &       & 6239.66 &  0.02 &        \\
 \eu{5p}{2}{P}{\circ}{} - \eu{6d}{2}{D}{}{} & 12.88 & 6818.41 & -0.52 &  7.94  \\   
                                            &       & 6829.80 & -0.26 &  7.87  \\
                                            &       & 6829.83 & -1.22 &        \\
 \eu{4d}{2}{D}{}{} - \eu{5f}{2}{F}{\circ}{} & 12.52 & 7848.82 &  0.32 &  8.36  \\
                                            &       & 7849.72 &  0.47 &  8.29  \\
                                            &       & 7849.62 & -0.83 &        \\
\eu{5p}{2}{P}{\circ}{} - \eu{7s}{2}{S}{}{}  & 12.88 & 7125.85 & -0.79 &  7.73  \\
\eu{5f}{2}{F}{\circ}{} - \eu{9g}{2}{G}{}{}  & 14.10 & 7911.52 & -0.42 &  7.83  \\
                                            &       & 7911.63 & -0.60 &        \\
\eu{5d}{2}{D}{}{} - \eu{8f}{2}{F}{\circ}{}  & 13.94 & 8044.41 & -0.59 & 7.58   \\
                                            &       & 8044.55 & -0.44 &        \\
\eu{5f}{2}{F}{\circ}{} - \eu{8g}{2}{G}{}{}  & 14.10 & 8935.50 & -0.12 & 7.94   \\
                                            &       & 8935.63 & -0.30 &        \\
\eu{4f}{2}{F}{\circ}{} - \eu{5g}{2}{G}{}{}  & 12.84 & 9412.66 &  1.23 & 7.96   \\
                                            &       & 9412.66 & -0.31 &        \\
                                            &       & 9412.78 &  1.12 &        \\
\noalign{\smallskip}\hline \noalign{\smallskip}
\multicolumn{5}{l}{{\bf Notes.} See text for sources of $gf$-values.}
	\end{tabular}     
\end{table}              

\begin{figure}
	% To include a figure from a file named example.*
	% Allowable file formats are eps or ps if compiling using latex
	% or pdf, png, jpg if compiling using pdflatex
	\includegraphics[width=\columnwidth]{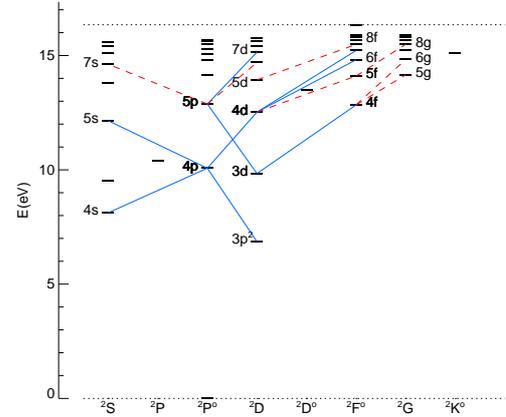}
    \caption{The doublet terms in the model atom of Si\ii. The absorption and emission spectral lines observed  in $\iota$~Her arise from the transitions shown as continuous and dashed lines, respectively.}
    \label{fig:si2_term}
\end{figure}

\begin{figure*}           
 \begin{minipage}{150mm}

\hspace{-10mm}
\parbox{0.3\linewidth}{\includegraphics[scale=0.55]{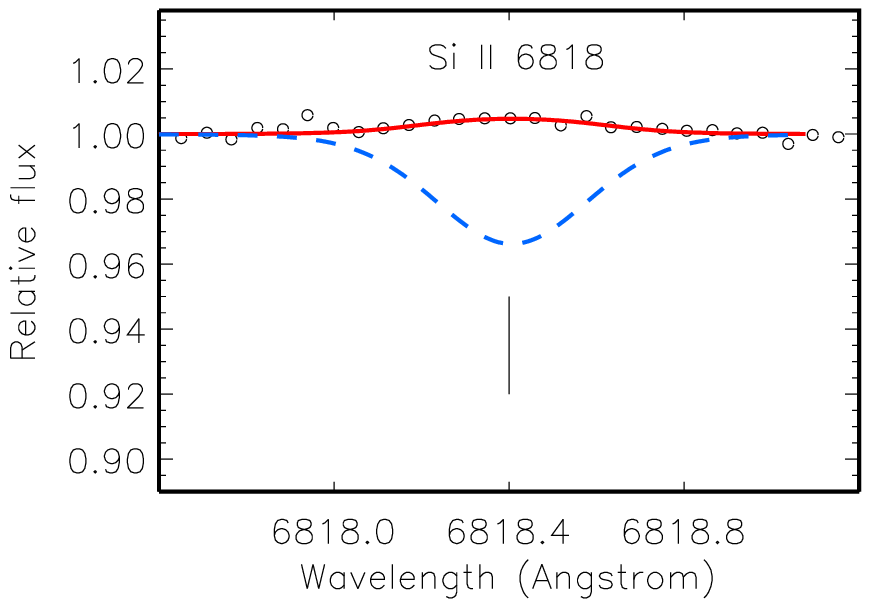}\\
\centering}
%\hspace{0.2\linewidth}
\hspace{5mm}
\parbox{0.3\linewidth}{\includegraphics[scale=0.55]{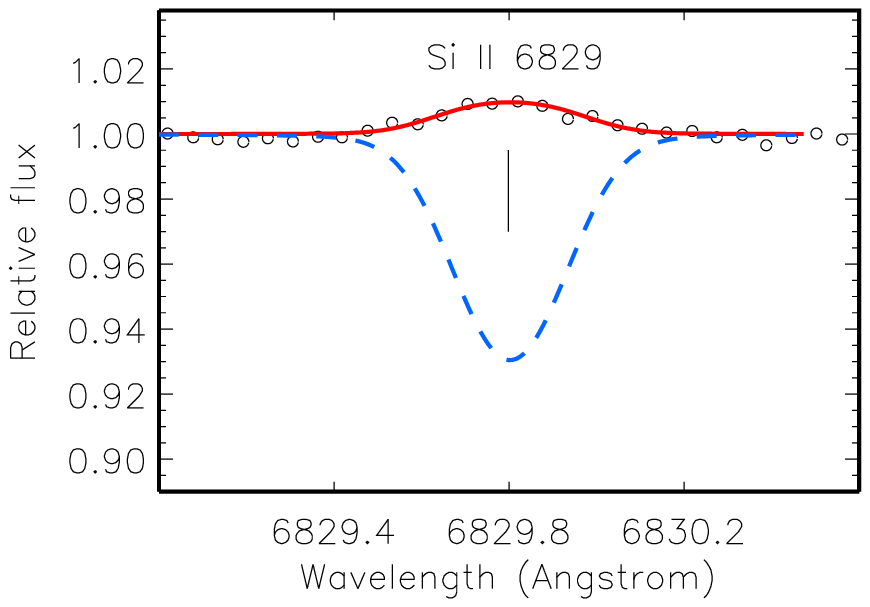}\\
\centering}
\hspace{5mm}
\parbox{0.3\linewidth}{\includegraphics[scale=0.55]{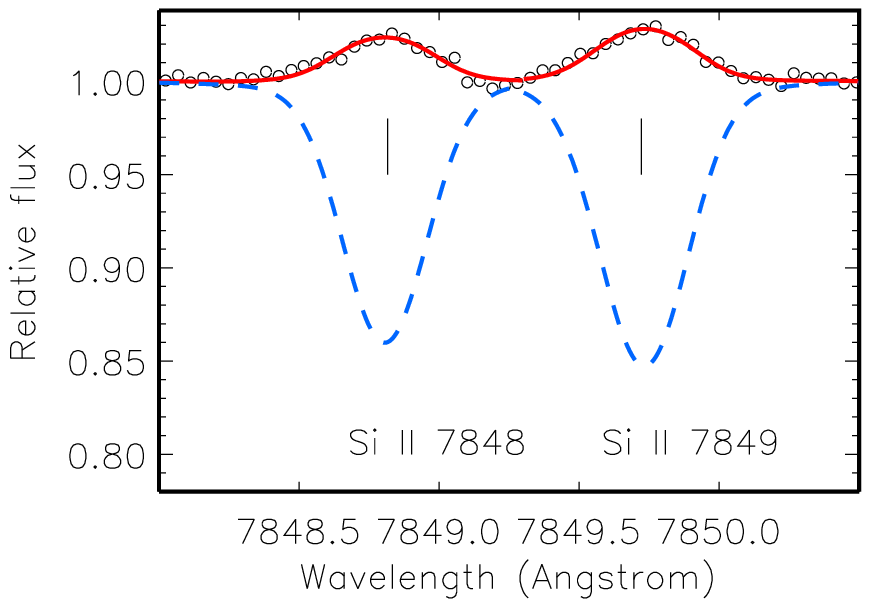}\\
\centering}
%\hspace{1\linewidth}
\hfill
\\[0ex]

\hspace{-10mm}
\parbox{0.3\linewidth}{\includegraphics[scale=0.55]{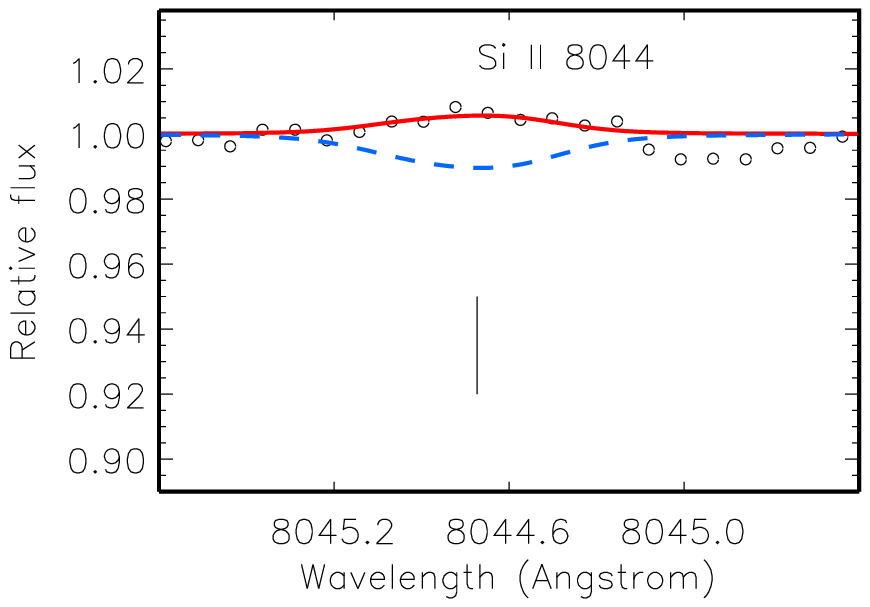}\\
\centering}
%\hspace{0.2\linewidth}
\hspace{5mm}
\parbox{0.3\linewidth}{\includegraphics[scale=0.55]{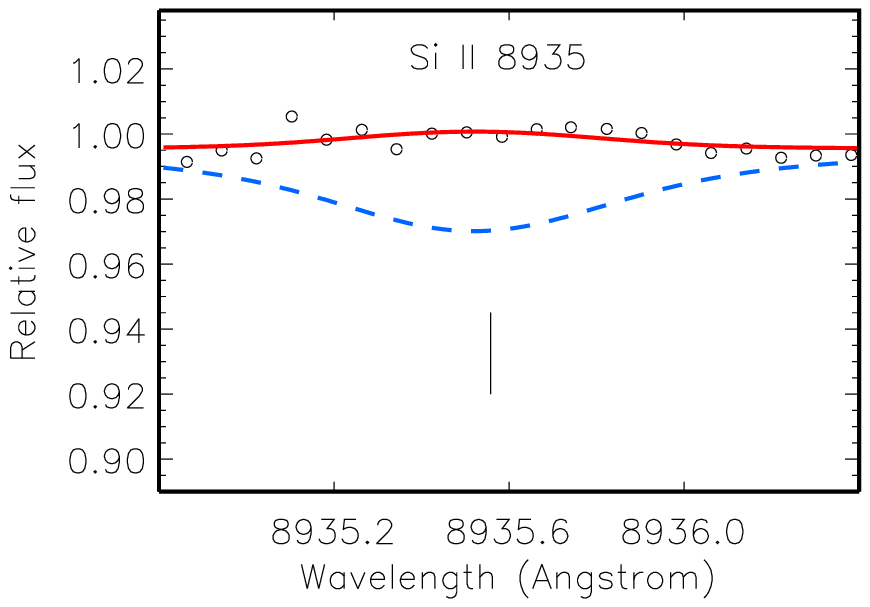}\\
\centering}
\hspace{5mm}
\parbox{0.3\linewidth}{\includegraphics[scale=0.55]{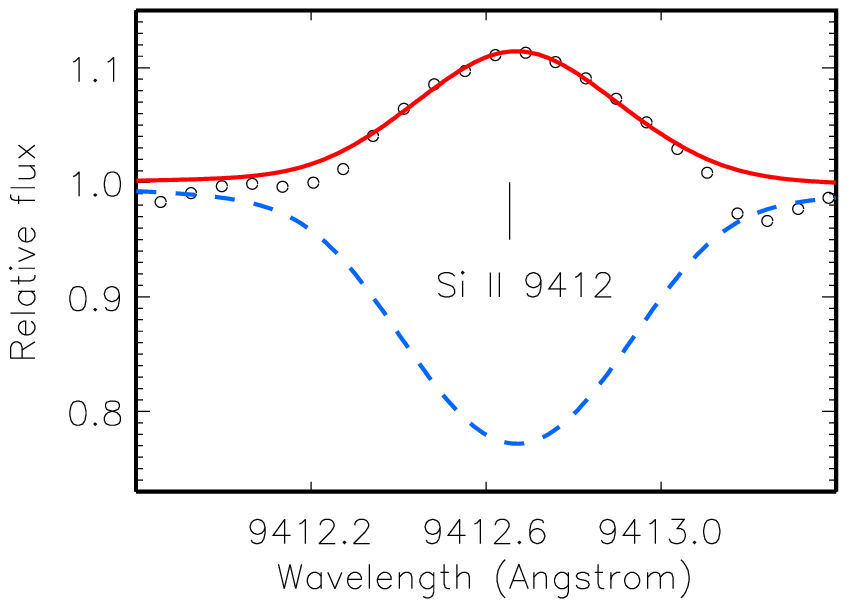}\\
\centering}
%\hspace{1\linewidth}
\hfill
\\[0ex]
\caption{Best NLTE fits (solid curves) of the emission Si\ii\ lines in $\iota$~Her (open circles). The obtained Si abundances are presented in Table~\ref{tab:abund_stars}. The corresponding LTE line profiles are shown by dashed curves. 
%For Si\ii\ 5688 and 8935\,\AA, the calculations were performed with $\eps{}$ = 7.50 and 7.94, respectively.
}
\label{si2_emission}
\end{minipage}
\end{figure*} 

\section{NLTE analysis of $\iota$~Her}\label{sect:iher}

\subsection{Silicon absorption lines}

%\underline{$\iota$~Her.} Here, we do not consider the Si\ii\ emission lines, which are discussed in the next section. 
NLTE reduces $\sigma$ of the Si\ii\ based abundance, however, the line-to-line scatter remains uncomfortably large (Fig.~\ref{fig:hd72660_picet}). It is largely contributed by lines of the two multiplets, 3853-62\,\AA\ (\Eexc\ = 6.86~eV) and 6347-71\,\AA\ (\Eexc\ = 8.12~eV). For example, for Si\ii\ in NLTE, we obtained $\sigma$ = 0.50~dex, when taking Si\ii\ 6347-71\,\AA\ into account, and $\sigma$ = 0.25~dex without these two lines. 
%The situation is even worse in LTE: $\sigma$ = 0.85 and 0.41~dex, respectively. 
In order to be consistent with the other lines, Si\ii\ 6347 and 6371\,\AA\ need to be strengthened in NLTE like that shown in Fig.~\ref{fig:si6371} for a model, where the TOPbase photoionisation cross sections were replaced with the hydrogenic ones for the Si\ii\ $4s$, $3d$, and \eu{3p^2}{2}{D}{}{} levels. Such a change in the NLTE effects compared with that for our standard model atom can be understood. From inspecting the NET = $n_l R_{lu} - n_u R_{ul}$ rates for b-f transitions of Si\ii, it was found that overionisation of exactly these three levels is the main driver of overal overionisation of Si\ii. Here, $R_{lu}$ and $R_{ul}$ are radiative rates of the $l-u$ and the reverse transitions, respectively.
%An increase of radiative rates for the pumping transitions from \eu{3p}{2}{P}{\circ}{} to \eu{4s}{2}{S}{}{} (1526, 1533\,\AA), \eu{3p^2}{2}{S}{}{} (1304, 1309\,\AA), and \eu{3p^2}{2}{D}{}{} (1808, 1816\,\AA), by one order of magnitude, makes Si\ii\ 6347, 6371\,\AA\ stronger and reduces the derived abundance by about 0.3~dex. This is still not enough to  
%or replacement of electron-impact excitation data of \citet{2014MNRAS.442..388A} with the \citet{Reg1962} semi-empirical approximation Applying hydrogenic photoionisation cross sections instead of the TOPBASE ones (Fig.~\ref{fig:si6371}) could reduce the abundance derived from Si\ii\ 6347, 6371\,\AA, 
For the Si\ii\ $4s$ and $3d$ levels, their hydrogenic photoionisation cross sections are smaller than the TOPbase ones, by more than one order of magnitude near the thresholds. Therefore, their use results in weakened NLTE effects for Si\ii\ and fits better to Si\ii\ 6347, 6371\,\AA. However, with such an atomic model, we cannot reproduce the Si\ii\ emission lines (see Sect.~\ref{sect:emission}). 

Even without these two lines, the Si\ii\ based abundance is higher than that from the Si\iii\ lines, by 0.20~dex. For comparison, under the LTE assumption, an abundance difference between Si\ii\ and Si\iii\ amounts to $-0.72$~dex. 
When relying on the Si\iii\ lines, close-to-solar silicon abundance of [Si/H] = 0.03$\pm$0.07 was obtained for $\iota$~Her, in line with the earlier determinations of \citet{2012A&A...539A.143N}. 

\subsection{Uncertainties in derived abundances}

We checked a sensitivity of the NLTE abundances from representative lines of Si\ii\ and Si\iii\ to variations in atomic model and atmospheric parameters. Table~\ref{tab:uncertainties} summarises results of our tests. 

The NLTE effects for Si\ii\ 6347, 6371\,\AA\ were found to be rather stable. Variations in photoionisation cross sections and collisional rates produce abundance shifts of up to 0.07~dex. A substantial reduction of $\Delta_{\rm NLTE}$, by 0.33~dex, was obtained in the only case, where oscillator strengths of the pumping \eu{3p}{2}{P}{\circ}{} -- \eu{3p^2}{2}{D}{}{} (1808-17~\AA), \eu{3p^2}{2}{S}{}{} (1304-09~\AA), and \eu{4s}{2}{S}{}{} (1526-33~\AA) transitions were increased 
%The NLTE abundance from these lines can be substantially reduced, by 0.33~dex, only when increasing oscillator strengths of the pumping \eu{3p}{2}{P}{\circ}{} -- \eu{3p^2}{2}{D}{}{}, \eu{3p^2}{2}{S}{}{}, and \eu{4s}{2}{S}{}{} transitions, 
by a factor of 10. Such a big uncertainty in $f_{lu}$ is unlikely.
%resulted in abundance change of no more than 0.1~dex for Si\ii\ 6347, 6371\,\AA. 
We shall conclude that we are not able to treat correctly the formation of Si\ii\ 6347, 6371\,\AA\ in $\iota$~Her. These two lines were not accounted for, when computing the mean abundance from the Si\ii\ lines. 

\citet{2012A&A...539A.143N} estimate the uncertainties in $\Teff$, $\logg$, and $\xi_t$ as 200~K, 0.05~dex, and 1~\kms, respectively.
% Table~\ref{tab:uncertainties} shows effects of variations in atmospheric parameters on derived abundances for the representative lines. 
A downward revision of $\Teff$ would reduce the mean abundance from lines of Si\ii, by 0.03~dex, and, in contrast, would increase the mean abundance for Si\iii, by 0.08~dex, removing, in part, a discrepancy between Si\ii\ and Si\iii. Upward revision of $\logg$ would act in the same direction, however, the abundance shifts are smaller. 

For the majority of the Si\ii\ lines in $\iota$~Her, we use the $\Gamma_4/N_{\rm e}$ values from laboratory measurements of \citet{2009A&A...508..491B}. They are proven to work well for the stars (Sect.~\ref{sect:mild}), where the quadratic Stark effect broadening is stronger than for $\iota$~Her. While approximate formula of \citet{1971Obs....91..139C} was applied for lines of Si\iii. A reduction of $\Gamma_4$ by a factor of two would increase the Si\iii\ based abundance, by 0.05~dex (Table~\ref{tab:uncertainties}). 

Thus, the obtained abundance difference between Si\ii\ (without 6347, 6371\,\AA) and Si\iii\ can arise due to the uncertainties in atmospheric parameters and line data. 

% leads to a big abundance discrepancy of  , which is entirely removed in NLTE, so that the final Si\ii\ based abundance is even 
%Since the Si\ii\ abundance has large statistical error, 

\subsection{Origin of Si\ii\ emission lines}\label{sect:emission}

Ten lines of Si\ii, which arise from the transitions between the high-excitation doublet terms (Fig.\,\ref{fig:si2_term}), reveal emission or disappear in spectrum of $\iota$~Her. They are listed in Table~\ref{tab:emission_si2} and, in part, are displayed in Figs.~\ref{fig:si6239} and \ref{si2_emission}. Their $gf$-values are based on calculations of \citet{1995JPhB...28.3485M}, as presented by NIST. An exception is the Si\ii\ 9412\,\AA\ triplet. NIST provides log~$gf$ = $-0.306$ for the only line, which arises from the transition between the $J_{low} = 7/2$ and $J_{up} = 7/2$ sublevels. In order to obtain $gf$-values given in Table~\ref{tab:emission_si2}, we used calculations of R. Kurucz for all three lines of this triplet, as available in VALD: log~$gf$ = 1.012 (7/2 -- 9/2), $-0.532$ (7/2 -- 7/2), and 0.899 (5/2 -- 7/2) and applied their $gf$ ratios to the known NIST value.

The detected emission features are of photospheric origin and naturally explained by interlocked NLTE effects acting in a photosphere, like to emission lines of C\ione\ \citep{2016MNRAS.462.1123A} and Ca\ii\ \citep{2018MNRAS.477.3343S} in $\iota$~Her.
%8912-27, 9890
%as the line profiles are resolved and symmetric and  are  not  Doppler-shifted  relative  to  the  photospheric  absorption  lines. 
The main driving mechanism of emission is overionisation of Si\ii. As shown in Fig.~\ref{fig:bfactors} (bottom panel), the lower level of each emission transition is depopulated to the greater extent than is the upper level in the line-formation layers, resulting in a line source function which rises relative to the local Planck function with photospheric height. It is essential that the NLTE correction to stimulated emission is sensitive to small deviations   of the departure coefficients from unity whenever $h\nu \ll kT$. This explains why in the transitions from common lower level, $4d$, the shorter wavelength lines, at 4621 and 5466\,\AA, appear in absorption, although being greatly weakened compared with their LTE strengths, while the near-IR lines at 7848, 7849\,\AA\ are in emission.

We checked how the emission mechanism is stable with respect to a variation in atomic data. Table~\ref{tab:uncertainties} shows the abundance shifts for Si\ii\ 6239\,\AA\ in a part of all the tests made. Increasing photoionisation cross sections within their uncertainties (10~\%\ for the TOPbase data) and even twice has minor effect on the emission phenomenon. The effect is strong, so that emission in the Si\ii\ lines disappear, when the TOPbase photoionisation cross sections are either reduced by a factor of 10 or replaced, in part (for the Si\ii\ $4s$, $3d$, and \eu{3p^2}{2}{D}{}{} levels, Fig.\,\ref{fig:si6239}) with the hydrogenic ones. 

Another set of test calculations was performed by varying collisional recipes:
%Calculations of \citet{2014MNRAS.442..388A} do not include most of the transitions between levels above \eu{5p}{2}{P}{\circ}{}. Therefore, the calculations were performed with various collisional recipes concerning the high-excitation transitions: 
(i) electron-impact excitation data of \citet{2014MNRAS.442..388A} were replaced with the \citet{Reg1962} semi-empirical approximation for the allowed transitions and $\Upsilon$ = 1 for the forbidden transitions (Fig.\,\ref{fig:si6239} for Si\ii\ 6239\,\AA); (ii) and (iii) the electron-impact ionisation rates were scaled by a factor of 0.1 and 10; (iv) for each b-b transition, its collisional rate was scaled by a factor of 2; (v), (vi), and (vii) for the transitions missing in \citet{2014MNRAS.442..388A}, the \citet{Reg1962} rates and the standard $\Upsilon$ = 1 value were scaled by a factor of 0.1, 0.5, and 2. Results of the tests (iv), (vi), and (vii) for Si\ii\ 6239\,\AA\ are given in Table~\ref{tab:uncertainties}. Emission in Si\ii\ 6239\,\AA\ disappear, if either to decrease the electron-impact ionisation rates (ii), or to increase the electron-impact excitation rates (iv, vii). The remaining tests keep emission in the Si\ii\ lines, although the best line profile fits are achieved with different element abundances for different collisional recipes. 
%A reduction of collisional rates by a factor of 10 for the forbidden transitions between high-excitation levels (vi) makes absorption for all lines of Si\ii.

%In test calculations, their TOPBASE photoionisation cross sections were replaced with the hydrogenic ones.
Using our standard model atom, we derived the NLTE abundances from fitting the emission line profiles (Table~\ref{tab:emission_si2} and Fig.~\ref{fig:hd72660_picet}). They lie between $\eps{}$ = 7.58, which is close to the Si\iii\ based abundance, and substantially higher value of $\eps{}$ = 8.36. Our test calculations show that a magnitude of emission is very sensitive to a variation in collisional rates for the high-excitation transitions. Exactly those transitions are missing in calculations of \citet{2014MNRAS.442..388A} that include the levels below \eu{5p}{2}{P}{\circ}{}. Extended calculations of electron-impact excitation cross sections for Si\ii\ are highly desirable for achieving consistent abundances from different emission lines.
%In order to make the range of abundances derived from the emission lines narrower, we performed many , by varying radiative and collisional data for b-b and b-f transitions of Si\ii, however, the solution was not found.

%The  essence  of  the  emission  mechanism  is  the  sensitivity of the non-LTE correction to stimulated emission to small deviations   of   the   departure   coefficients   from   unity   whenever.
%non-LTE effects which decouple the line source function from and 

\citet{2019PASJ...71...45S} registered 12 emission lines of Si\ii\ in $\iota$~Her. Our NLTE calculations reproduce ten of them (Table~\ref{tab:emission_si2}). For the remaining two lines, at 5688 and 5701\,\AA, the NLTE calculations predict too strong emission. In order to fit to the observations, one needs to reduce the Si abundance compared with the solar value, by more than one order of magnitude. Both lines arise from the transition \eu{3p3d}{4}{F}{\circ}{} -- \eu{3p4p}{4}{D}{}{}, with \Eexc\ $>$ 14~eV for the lower level. The majority of the quartet atomic terms have energies higher than the ionisation energy of the Si\ii\ ground state, so that our model atom includes a few number (seven) of quartet terms.
%, which lie below the ionisation threshold. 
These levels are weakly coupled to the doublet terms, mostly via collisional processes, and their populations are more dependent of the transitions between the quartet levels. Since the system of quartet terms in our model atom is certainly incomplete, we cannot compute their populations correctly. For example, our NLTE calculations predict emission for the multiplet 6665, 6671, 6699\,\AA\ (\eu{3p4s}{4}{P}{\circ}{} -- \eu{3p4p}{4}{D}{}{}), while these lines are observed in absorption.

%Test calculations
%increase fij by a factor of 10 for transitions from the ground state to 4s, 3d, 3p2 2D, 3p2 2S, 3p2 2P,
%forbidden: van Reg with fij = 0.1 for common n and fij = 0.001 for the remaining

\section{NLTE abundance corrections depending on atmospheric parameters}\label{sect:dnlte}

\begin{figure}
	% To include a figure from a file named example.*
	% Allowable file formats are eps or ps if compiling using latex
	% or pdf, png, jpg if compiling using pdflatex
	\includegraphics[width=\columnwidth]{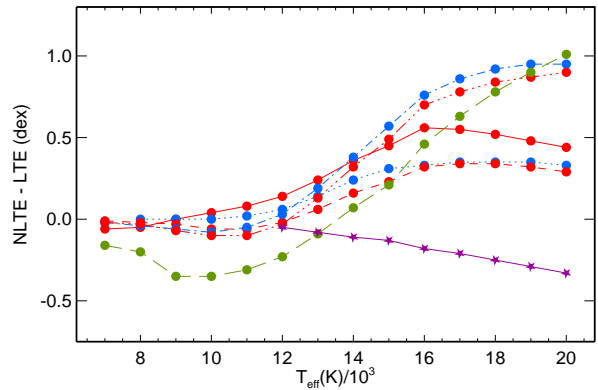}
    \caption{NLTE abundance corrections for Si\ii\ lines (circles) and Si\iii\ 4567~\AA\ (5 pointed stars) depending on $\Teff$ in the models with common $\logg$ = 4.0 and solar metallicity. Different type curves correspond to different lines of the Si\ii: 3862~\AA\ (solid), 4075~\AA\ (dotted), 4128~\AA\ (short-dashed), 5055~\AA\ (dot-dashed), 5978~\AA\ (three-dot-dashed), and 6371~\AA\ (long-dashed).} 
    \label{fig:dnlte}
\end{figure}

Our analyses of the silicon lines in the nine stars with well determined atmospheric parameters provide evidence for a correct treatment of the NLTE line formation for Si\ione -\ii -\iii\ through a range of A-B spectral types. Therefore, we can recommend the users to apply in their research the NLTE abundance corrections computed in this study. They are available for the lines listed in Tables~\ref{tab:sun_hd32115} (Si\ione), \ref{tab:abund_stars} (Si\ione\ 3905~\AA\ and Si\ii), and \ref{tab:abund_si3} (Si\iii). For calculations of $\Delta_{\rm NLTE}$s, the model atmospheres with solar metallicity and $\logg$ = 3.5, 4.0, and 4.5 were taken from the Kurucz's grid\footnote{http://kurucz.harvard.edu/grids/gridp00odfnew/}. For Si\ione, Si\ii, and Si\iii, $\Teff$ varies between 7000 and 9000~K, 7000 and 20000~K, and 12\,000 and 20000~K, respectively, with a step of 1000~K.
%Ranges of $\Teff$ were different for Si\ione, Si\ii, and Si\iii. For the Si\ione\ lines, except Si\ione\ 3905~\AA, $\Teff$ = 7\,000, 8\,000, and 9\,000~K, because their equivalent widths  decrease steeply with effective temperature.  For Si\iii, the calculations were performed in the 12\,000 $\le \Teff \le$ 20\,000~K range. 
Tables~\ref{tab:nlte_corr1} and \ref{tab:nlte_corr23} and Fig.~\ref{fig:dnlte} present, in part, the computed $\Delta_{\rm NLTE}$s. 

% were computed Equivalent widths of  

\begin{table}
	\centering
	\caption{NLTE abundance corrections (dex) for lines of Si\ione\ in the grid of model atmospheres.  }
	\label{tab:nlte_corr1} \tabcolsep5.2mm
	\begin{tabular}{crrr} % 
		\hline\hline \noalign{\smallskip}
$\lambda$ [\AA] & \multicolumn{3}{c}{$\Teff$ [kK], $\logg$ = 4.0} \\
		\cline{2-4} \noalign{\smallskip}
         &     7 &     8 &     9  \\
		\hline \noalign{\smallskip}
 5948 & $-$0.04 &  0.02 &  0.46 \\
 6155 & $-$0.03 & $-$0.00 &  0.16 \\
 7405 & $-$0.05 & $-$0.01 &  0.24 \\
\noalign{\smallskip}\hline \noalign{\smallskip}
\multicolumn{4}{l}{This table is available in its entirety in } \\
\multicolumn{4}{l}{a machine-readable form in the online journal. } \\
\multicolumn{4}{l}{A portion is shown here for guidance} \\
\multicolumn{4}{l}{ regarding its form and content. } \\
\noalign{\smallskip} \hline
	\end{tabular} 
\end{table}  

\begin{table*}
	\centering
	\caption{NLTE abundance corrections (dex) for lines of Si\ione, Si\ii, and Si\iii\ in the grid of model atmospheres.  }
	\label{tab:nlte_corr23}
	\begin{tabular}{crrrrrrrrrrrrrr} % 
		\hline\hline \noalign{\smallskip}
$\lambda$ [\AA] & \multicolumn{14}{c}{$\Teff$ [kK], $\logg$ = 4.0} \\
		\cline{2-15} \noalign{\smallskip}
         &     7 &     8 &     9 &    10 &    11 &    12 &    13 &    14 &    15 &    16 &    17 &    18 &    19 &    20 \\
		\hline \noalign{\smallskip}
Si\ione & \multicolumn{14}{l}{} \\
3905 & $-$0.01 &  0.03 &  0.34 &  0.43 &  0.36 &  0.39 &  0.42 &       &       &       &       &       &       &       \\
Si\ii & \multicolumn{14}{l}{} \\
5978 & $-$0.02 & $-$0.03 & $-$0.07 & $-$0.10 & $-$0.10 & $-$0.03 &  0.13 &  0.32 &  0.49 &  0.70 &  0.78 &  0.84 &  0.87 &  0.90 \\
6239 &       &       &  0.05 &  0.07 &  0.14 &  0.27 &  0.54 &  1.07 &  1.12 &  1.17 &       &  e    &  e    & e     \\
6371 & $-$0.16 & $-$0.20 & $-$0.35 & $-$0.35 & $-$0.31 & $-$0.23 & $-$0.09 &  0.07 &  0.21 &  0.46 &  0.63 &  0.78 &  0.90 &  1.01 \\
Si\iii & \multicolumn{14}{l}{} \\ 
4567 &       &       &       &       &       & $-$0.05 & $-$0.08 & $-$0.11 & $-$0.13 & $-$0.18 & $-$0.21 & $-$0.25 & $-$0.29 & $-$0.33 \\
\noalign{\smallskip}\hline \noalign{\smallskip}
\multicolumn{15}{l}{For a given line, $\Delta_{\rm NLTE}$ is provided, if the NLTE equivalent width exceeds 3~m\AA. e = emission line.} \\
\multicolumn{15}{l}{This table is available in its entirety in a machine-readable form in the online journal. A portion is shown here} \\
\multicolumn{15}{l}{ for guidance regarding its form and content. } \\
\noalign{\smallskip} \hline
	\end{tabular} 
\end{table*}  

For the lines lying in the wings of the hydrogen lines, that is, Si\ione\ 3905~\AA, Si\ii\ 3853, 3856, 3862, 4075, 4076, 4128, and 4130\,\AA, we recommend to determine abundances from the NLTE spectral synthesis method, but not simply adding the NLTE corrections to the LTE abundances. Our $\Delta_{\rm NLTE}$s for Si\ii\ 6347, 6371\,\AA\ should be applied with a caution in case of the stars as hot as $\iota$~Her (see Sect.~\ref{sect:emission}).

\section{Conclusions}\label{sect:Conclusions}

This paper presents a new comprehensive model atom of Si\ione -\ii -\iii\ that can be applied to the NLTE analyses of Si\ione, Si\ii, and Si\iii\ lines in a wide range of spectral types, where they are observed. Here, we performed the NLTE calculations for a range of atmospheric parameters characteristic  of unevolved A-B type stars: $\Teff$ = 7000 to 20\,000~K, $\logg$ = 3.5, 4.0, 4.5, and solar metallicity. The NLTE effects are different for lines of different ions in a given model atmosphere, and they depend strongly on $\Teff$.

For lines of Si\ione, the NLTE effects are small for $\Teff \le$ 8000~K, with slightly negative $\Delta_{\rm NLTE}$ at $\Teff$ = 7000~K and slightly positive ones at $\Teff$ = 8000~K. The \Eexc\ $>$ 4.9~eV lines are steeply weakened with increasing $\Teff$ and most of them cannot be measured in the $\Teff \sim$ 9000~K stars. For the strongest of them and Si\ione\ 3905~\AA, $\Delta_{\rm NLTE}$ grows towards higher $\Teff$ and reaches 0.42~dex for Si\ione\ 3905~\AA\ in the 13\,000/4.0 model atmosphere.

The first ion of silicon is a majority species in the line formation layers of the $\Teff \le$ 11\,000~K models and is subject to overionisation in the hotter atmospheres. Except for Si\ii\ 6347, 6371\,\AA, the NLTE effects are small for the Si\ii\ lines, where Si\ii\ is a majority species. For different lines, $\Delta_{\rm NLTE}$ can be of different sign in a given atmosphere and does not exceed 0.1~dex, in absolute value. The same lines in the $\Teff \ge$ 12\,000~K models are weakened compared with their LTE strengths due to overionisation of Si\ii, resulting in positive $\Delta_{\rm NLTE}$s, which grow with increasing $\Teff$. For the Si\ii\ 3853-62\,\AA, 4075-76\,\AA, and 4128-30\,\AA\ multiplets, $\Delta_{\rm NLTE}$s reach their maximal values at $\Teff$ of about 17\,000~K and decrease for higher $\Teff$ due to shifting the line formation depths to deeper atmospheric layers. The NLTE effects are, in particular, large for the Si\ii\ 5041-56\,\AA\ and 5957-78\,\AA\ multiplets, with $\Delta_{\rm NLTE}$ of up 0.95~dex in the 20\,000/4.0 model.

The Si\ii\ 6347, 6371\,\AA\ lines are strongly strengthened in NLTE over wide range of $\Teff$, even in the atmospheres, where Si\ii\ is subject to overionisation. 
For example, for Si\ii\ 6347\,\AA\ in the $\logg$ = 4.0 models, $\Delta_{\rm NLTE}$ reaches the most negative value of $-0.40$~dex for $\Teff$ = 9000 and 10\,000~K and turns to positive for $\Teff \ge$ 14\,000~K.

Our NLTE calculations predict that, in the $\Teff \ge$ 17\,000~K models, some lines of Si\ii, which arise from the high-excitation (above \eu{4d}{2}{D}{}{}) doublet levels, come into emission due to the NLTE effects acting in an atmosphere. The main driving mechanism of emission is overionisation of Si\ii.

NLTE abundance corrections for the Si\iii\ lines are negative in the stellar parameter range, with which we concern, and they increase, in absolute value, toward higher $\Teff$.

The new model atom was tested with nine unevolved A9 to B3-type stars, with well determined atmospheric parameters and high-resolution observed spectra available. The hottest star, $\iota$~Her, is known by many emission lines of various chemical species, including 12 lines of Si\ii\ \citep{2019PASJ...71...45S}. Our NLTE calculations with a classical hydrostatic model that represents the atmosphere of $\iota$~Her reproduced ten of them, although using rather different element abundances for different lines. A magnitude of emission is sensitive to a variation in collisional rates.  Accurate electron-impact excitation cross sections for the Si\ii\ transitions between the high-excitation leves are highly desirable, in order to achieve consistent abundances from different emission lines.

For each star, the NLTE and LTE abundances were determined from the absorption lines. NLTE reduces substantially the line-to-line scatter for Si\ii\ compared with the LTE case and leads to consistent mean abundances from lines of different ionisation stages. 
For example, in NLTE and LTE, the Si\ione\ -- Si\ii\ abundance difference in 21~Peg amounts to $-0.01$ and $-0.54$~dex, respectively, and the Si\ii\ -- Si\iii\ difference in $\iota$~Her is +0.20 and $-0.72$~dex. 
Thus, with the new model atom, the line formation for Si\ione\ - Si\ii\ - Si\iii\ in atmospheres of A to mid B-type stars is treated correctly. The exception is the Si\ii\ 6347, 6371\,\AA\ doublet in $\iota$~Her, for which the NLTE effects are overestimated. They are stable with respect to variations in photoionisation cross sections and electron-impact excitation data. At this stage, we failed to understand this problem and further theoretical work is required.

We obtained that, except for $\pi$~Cet, the silicon abundance follows the iron one in our sample stars, including a $\lambda$~Boo star Vega and Am stars HD~72660 and Sirius. This suggests that the mechanisms, which produced deviations in metal abundances of our chemically peculiar stars from the solar one, did not separate chemical elements. A supersolar abundance of [Si/H] = 0.23 was found from lines of the two ionisation stages, Si\ii\ and Si\iii, for $\pi$~Cet, which has [Fe/H] = 0, according to \citet{2009AA...503..945F}. A NLTE study of the iron lines in A to mid B-type stars is in progress (Sitnova et~al., in prep.)

%   \software{DETAIL \citep{detail}, SynthV\_NLTE \citep{2019ASPC}, binmag \citep{binmag3,2018ascl.soft05015K}, LLmodels code \citep{2004AA...428..993S}}.

{\it Acknowledgments.} 
The author thanks K.M.~Aggarwal for providing effective collision strengths for the Si\ii\ transitions, T.~Ryabchikova for providing the observed spectra and the model atmospheres computed with the code \textsc{LLmodels}, and J.~Landstreet for providing the UV spectrum of HD~72660.
 This study made use of the ESO UVESPOP, NIST, TOPbase, NORAD, open-ADAS\footnote{http://open.adas.
ac.uk}, VALD, ADS\footnote{http://adsabs.harvard.edu/abstract\_service.html}, and R.~Kurucz's databases.

\bibliography{atomic_data,mashonkina,nlte,si2018}

\begin{thebibliography}{}
\makeatletter
\relax
\def\mn@urlcharsother{\let\do\@makeother \do\$\do\&\do\#\do\^\do\_\do\%\do\~}
\def\mn@doi{\begingroup\mn@urlcharsother \@ifnextchar [ {\mn@doi@}
  {\mn@doi@[]}}
\def\mn@doi@[#1]#2{\def\@tempa{#1}\ifx\@tempa\@empty \href
  {http://dx.doi.org/#2} {doi:#2}\else \href {http://dx.doi.org/#2} {#1}\fi
  \endgroup}
\def\mn@eprint#1#2{\mn@eprint@#1:#2::\@nil}
\def\mn@eprint@arXiv#1{\href {http://arxiv.org/abs/#1} {{\tt arXiv:#1}}}
\def\mn@eprint@dblp#1{\href {http://dblp.uni-trier.de/rec/bibtex/#1.xml}
  {dblp:#1}}
\def\mn@eprint@#1:#2:#3:#4\@nil{\def\@tempa {#1}\def\@tempb {#2}\def\@tempc
  {#3}\ifx \@tempc \@empty \let \@tempc \@tempb \let \@tempb \@tempa \fi \ifx
  \@tempb \@empty \def\@tempb {arXiv}\fi \@ifundefined
  {mn@eprint@\@tempb}{\@tempb:\@tempc}{\expandafter \expandafter \csname
  mn@eprint@\@tempb\endcsname \expandafter{\@tempc}}}

\bibitem[\protect\citeauthoryear{{Aggarwal} \& {Keenan}}{{Aggarwal} \&
  {Keenan}}{2014}]{2014MNRAS.442..388A}
{Aggarwal} K.~M.,  {Keenan} F.~P.,  2014, \mn@doi [\mnras]
  {10.1093/mnras/stu883}, \href
  {http://adsabs.harvard.edu/abs/2014MNRAS.442..388A} {442, 388}

\bibitem[\protect\citeauthoryear{{Alexeeva}, {Ryabchikova}  \&
  {Mashonkina}}{{Alexeeva} et~al.}{2016}]{2016MNRAS.462.1123A}
{Alexeeva} S.~A.,  {Ryabchikova} T.~A.,   {Mashonkina} L.~I.,  2016, \mn@doi
  [\mnras] {10.1093/mnras/stw1635}, \href
  {http://adsabs.harvard.edu/abs/2016MNRAS.462.1123A} {462, 1123}

\bibitem[\protect\citeauthoryear{{Alexeeva}, {Ryabchikova}, {Mashonkina}  \&
  {Hu}}{{Alexeeva} et~al.}{2018}]{2018ApJ...866..153A}
{Alexeeva} S.,  {Ryabchikova} T.,  {Mashonkina} L.,   {Hu} S.,  2018, \mn@doi
  [\apj] {10.3847/1538-4357/aae1a8}, \href
  {https://ui.adsabs.harvard.edu/abs/2018ApJ...866..153A} {866, 153}

\bibitem[\protect\citeauthoryear{{Amarsi} \& {Asplund}}{{Amarsi} \&
  {Asplund}}{2017}]{2017MNRAS.464..264A}
{Amarsi} A.~M.,  {Asplund} M.,  2017, \mn@doi [\mnras] {10.1093/mnras/stw2445},
  \href {http://adsabs.harvard.edu/abs/2017MNRAS.464..264A} {464, 264}

\bibitem[\protect\citeauthoryear{{Bailey} \& {Landstreet}}{{Bailey} \&
  {Landstreet}}{2013}]{2013A&A...551A..30B}
{Bailey} J.~D.,  {Landstreet} J.~D.,  2013, \mn@doi [\aap]
  {10.1051/0004-6361/201220671}, \href
  {http://adsabs.harvard.edu/abs/2013A%26A...551A..30B} {551, A30}

\bibitem[\protect\citeauthoryear{{Bard} \& {Carlsson}}{{Bard} \&
  {Carlsson}}{2008}]{2008ApJ...682.1376B}
{Bard} S.,  {Carlsson} M.,  2008, \mn@doi [\apj] {10.1086/589910}, \href
  {http://adsabs.harvard.edu/abs/2008ApJ...682.1376B} {682, 1376}

\bibitem[\protect\citeauthoryear{{Becker} \& {Butler}}{{Becker} \&
  {Butler}}{1990}]{1990A&A...235..326B}
{Becker} S.~R.,  {Butler} K.,  1990, \aap, \href
  {http://adsabs.harvard.edu/abs/1990A%26A...235..326B} {235, 326}

\bibitem[\protect\citeauthoryear{{Bergemann}, {Kudritzki}, {W{\"u}rl}, {Plez},
  {Davies}  \& {Gazak}}{{Bergemann} et~al.}{2013}]{2013ApJ...764..115B}
{Bergemann} M.,  {Kudritzki} R.-P.,  {W{\"u}rl} M.,  {Plez} B.,  {Davies} B.,
  {Gazak} Z.,  2013, \mn@doi [\apj] {10.1088/0004-637X/764/2/115}, \href
  {https://ui.adsabs.harvard.edu/abs/2013ApJ...764..115B} {764, 115}

\bibitem[\protect\citeauthoryear{{Blanco}, {Botho}  \& {Campos}}{{Blanco}
  et~al.}{1995}]{1995PhyS...52..628B}
{Blanco} F.,  {Botho} B.,   {Campos} J.,  1995, \mn@doi [\physscr]
  {10.1088/0031-8949/52/6/004}, \href
  {https://ui.adsabs.harvard.edu/abs/1995PhyS...52..628B} {52, 628}

\bibitem[\protect\citeauthoryear{{Bukvi{\'c}}, {Djeni{\v{z}}e}  \&
  {Sre{\'c}kovi{\'c}}}{{Bukvi{\'c}} et~al.}{2009}]{2009A&A...508..491B}
{Bukvi{\'c}} S.,  {Djeni{\v{z}}e} S.,   {Sre{\'c}kovi{\'c}} A.,  2009, \mn@doi
  [\aap] {10.1051/0004-6361/200912046}, \href
  {https://ui.adsabs.harvard.edu/abs/2009A&A...508..491B} {508, 491}

\bibitem[\protect\citeauthoryear{{Butler} \& {Giddings}}{{Butler} \&
  {Giddings}}{1985}]{detail}
{Butler} K.,  {Giddings} J.,  1985, Newsletter on the analysis of astronomical
  spectra, No. 9, University of London

\bibitem[\protect\citeauthoryear{{Castelli} \& {Kurucz}}{{Castelli} \&
  {Kurucz}}{1993}]{1993ASPC...44..496C}
{Castelli} F.,  {Kurucz} R.~L.,  1993, in {Dworetsky} M.~M.,  {Castelli} F.,
  {Faraggiana} R.,  eds,  Astronomical Society of the Pacific Conference Series
  Vol. 44, IAU Colloq. 138: Peculiar versus Normal Phenomena in A-type and
  Related Stars. p.~496

\bibitem[\protect\citeauthoryear{{Cowley}}{{Cowley}}{1971}]{1971Obs....91..139C}
{Cowley} C.~R.,  1971, The Observatory, \href
  {https://ui.adsabs.harvard.edu/abs/1971Obs....91..139C} {91, 139}

\bibitem[\protect\citeauthoryear{{Cunto}, {Mendoza}, {Ochsenbein}  \&
  {Zeippen}}{{Cunto} et~al.}{1993}]{1993A&A...275L...5C}
{Cunto} W.,  {Mendoza} C.,  {Ochsenbein} F.,   {Zeippen} C.~J.,  1993, \aap,
  \href {http://adsabs.harvard.edu/abs/1993A%26A...275L...5C} {275, L5}

\bibitem[\protect\citeauthoryear{{Fern{\'a}ndez-Menchero}, {Del Zanna}  \&
  {Badnell}}{{Fern{\'a}ndez-Menchero} et~al.}{2014}]{2014A&A...572A.115F}
{Fern{\'a}ndez-Menchero} L.,  {Del Zanna} G.,   {Badnell} N.~R.,  2014, \mn@doi
  [\aap] {10.1051/0004-6361/201424849}, \href
  {https://ui.adsabs.harvard.edu/abs/2014A&A...572A.115F} {572, A115}

\bibitem[\protect\citeauthoryear{{Finn} \& {McAllister}}{{Finn} \&
  {McAllister}}{1978}]{1978SoPh...56..263F}
{Finn} G.~D.,  {McAllister} H.~C.,  1978, \mn@doi [\solphys]
  {10.1007/BF00152471}, \href
  {https://ui.adsabs.harvard.edu/abs/1978SoPh...56..263F} {56, 263}

\bibitem[\protect\citeauthoryear{{Fossati}, {Bagnulo}, {Monier}, {Khan},
  {Kochukhov}, {Landstreet}, {Wade}  \& {Weiss}}{{Fossati}
  et~al.}{2007}]{2007AA...476..911F}
{Fossati} L.,  {Bagnulo} S.,  {Monier} R.,  {Khan} S.~A.,  {Kochukhov} O.,
  {Landstreet} J.,  {Wade} G.,   {Weiss} W.,  2007, \mn@doi [\aap]
  {10.1051/0004-6361:20078320}, \href
  {http://adsabs.harvard.edu/abs/2007A%26A...476..911F} {476, 911}

\bibitem[\protect\citeauthoryear{{Fossati}, {Ryabchikova}, {Bagnulo},
  {Alecian}, {Grunhut}, {Kochukhov}  \& {Wade}}{{Fossati}
  et~al.}{2009}]{2009AA...503..945F}
{Fossati} L.,  {Ryabchikova} T.,  {Bagnulo} S.,  {Alecian} E.,  {Grunhut} J.,
  {Kochukhov} O.,   {Wade} G.,  2009, \mn@doi [\aap]
  {10.1051/0004-6361/200811561}, \href
  {http://adsabs.harvard.edu/abs/2009A%26A...503..945F} {503, 945}

\bibitem[\protect\citeauthoryear{{Fossati}, {Ryabchikova}, {Shulyak},
  {Haswell}, {Elmasli}, {Pandey}, {Barnes}  \& {Zwintz}}{{Fossati}
  et~al.}{2011}]{2011MNRAS.417..495F}
{Fossati} L.,  {Ryabchikova} T.,  {Shulyak} D.~V.,  {Haswell} C.~A.,  {Elmasli}
  A.,  {Pandey} C.~P.,  {Barnes} T.~G.,   {Zwintz} K.,  2011, \mn@doi [\mnras]
  {10.1111/j.1365-2966.2011.19289.x}, \href
  {http://adsabs.harvard.edu/abs/2011MNRAS.417..495F} {417, 495}

\bibitem[\protect\citeauthoryear{{Garz}}{{Garz}}{1973}]{1973A&A....26..471G}
{Garz} T.,  1973, \aap, \href
  {https://ui.adsabs.harvard.edu/abs/1973A&A....26..471G} {26, 471}

\bibitem[\protect\citeauthoryear{{Golriz} \& {Landstreet}}{{Golriz} \&
  {Landstreet}}{2016}]{2016MNRAS.456.3318G}
{Golriz} S.~S.,  {Landstreet} J.~D.,  2016, \mn@doi [\mnras]
  {10.1093/mnras/stv2658}, \href
  {http://adsabs.harvard.edu/abs/2016MNRAS.456.3318G} {456, 3318}

\bibitem[\protect\citeauthoryear{Gustafsson, Edvardsson, Eriksson, Jorgensen,
  Nordlund  \& Plez}{Gustafsson et~al.}{2008}]{Gustafssonetal:2008}
Gustafsson B.,  Edvardsson B.,  Eriksson K.,  Jorgensen U.~G.,  Nordlund
  {\AA}.,   Plez B.,  2008, A\&A, 486, 951

\bibitem[\protect\citeauthoryear{{Hill} \& {Landstreet}}{{Hill} \&
  {Landstreet}}{1993}]{1993AA...276..142H}
{Hill} G.~M.,  {Landstreet} J.~D.,  1993, \aap, \href
  {http://adsabs.harvard.edu/abs/1993A%26A...276..142H} {276, 142}

\bibitem[\protect\citeauthoryear{{Hubeny}, {Hummer}  \& {Lanz}}{{Hubeny}
  et~al.}{1994}]{1994A&A...282..151H}
{Hubeny} I.,  {Hummer} D.~G.,   {Lanz} T.,  1994, \aap, \href
  {https://ui.adsabs.harvard.edu/abs/1994A&A...282..151H} {282, 151}

\bibitem[\protect\citeauthoryear{{Hummer} \& {Mihalas}}{{Hummer} \&
  {Mihalas}}{1988}]{1988ApJ...331..794H}
{Hummer} D.~G.,  {Mihalas} D.,  1988, \mn@doi [\apj] {10.1086/166600}, \href
  {https://ui.adsabs.harvard.edu/abs/1988ApJ...331..794H} {331, 794}

\bibitem[\protect\citeauthoryear{{Kamp}}{{Kamp}}{1978}]{1978ApJS...36..143K}
{Kamp} L.~W.,  1978, \mn@doi [\apjs] {10.1086/190494}, \href
  {https://ui.adsabs.harvard.edu/abs/1978ApJS...36..143K} {36, 143}

\bibitem[\protect\citeauthoryear{{Kramida}, {Ralchenko}, {Reader}  \&
  Team}{{Kramida} et~al.}{2019}]{NIST19}
{Kramida} A.,  {Ralchenko} Y.,  {Reader} J.,   Team N.~A.,  2019, NIST Atomic
  Spectra Database (version 5.7.1).
USA, \url {http://physics.nist.gov/asd}

\bibitem[\protect\citeauthoryear{{Kurucz}, {Furenlid}, {Brault}  \&
  {Testerman}}{{Kurucz} et~al.}{1984}]{Atlas}
{Kurucz} R.~L.,  {Furenlid} I.,  {Brault} J.,   {Testerman} L.,  1984, {Solar
  flux atlas from 296 to 1300 nm}.
New Mexico: National Solar Observatory

\bibitem[\protect\citeauthoryear{{Lennon}, {Brown}, {Dufton}  \&
  {Lynas-Gray}}{{Lennon} et~al.}{1986}]{1986MNRAS.222..719L}
{Lennon} D.~J.,  {Brown} P.~J.~F.,  {Dufton} P.~L.,   {Lynas-Gray} A.~E.,
  1986, \mn@doi [\mnras] {10.1093/mnras/222.4.719}, \href
  {https://ui.adsabs.harvard.edu/abs/1986MNRAS.222..719L} {222, 719}

\bibitem[\protect\citeauthoryear{{Lodders}}{{Lodders}}{2019}]{2019arXiv191200844L}
{Lodders} K.,  2019, arXiv e-prints, \href
  {https://ui.adsabs.harvard.edu/abs/2019arXiv191200844L} {p. arXiv:1912.00844}

\bibitem[\protect\citeauthoryear{{Martin} \& {Zalubas}}{{Martin} \&
  {Zalubas}}{1983}]{1983JPCRD..12..323M}
{Martin} W.~C.,  {Zalubas} R.,  1983, \mn@doi [Journal of Physical and Chemical
  Reference Data] {10.1063/1.555685}, \href
  {https://ui.adsabs.harvard.edu/abs/1983JPCRD..12..323M} {12, 323}

\bibitem[\protect\citeauthoryear{{Matheron}, {Escarguel}, {Redon}, {Lesage}  \&
  {Richou}}{{Matheron} et~al.}{2001}]{2001JQSRT..69..535M}
{Matheron} P.,  {Escarguel} A.,  {Redon} R.,  {Lesage} A.,   {Richou} J.,
  2001, \mn@doi [\jqsrt] {10.1016/S0022-4073(00)00087-X}, \href
  {https://ui.adsabs.harvard.edu/abs/2001JQSRT..69..535M} {69, 535}

\bibitem[\protect\citeauthoryear{{Mendoza}, {Eissner}, {LeDourneuf}  \&
  {Zeippen}}{{Mendoza} et~al.}{1995}]{1995JPhB...28.3485M}
{Mendoza} C.,  {Eissner} W.,  {LeDourneuf} M.,   {Zeippen} C.~J.,  1995,
  \mn@doi [Journal of Physics B Atomic Molecular Physics]
  {10.1088/0953-4075/28/16/006}, \href
  {https://ui.adsabs.harvard.edu/abs/1995JPhB...28.3485M} {28, 3485}

\bibitem[\protect\citeauthoryear{{Mihalas}, {Hummer}  \& {Conti}}{{Mihalas}
  et~al.}{1972}]{1972ApJ...175L..99M}
{Mihalas} D.,  {Hummer} D.~G.,   {Conti} P.~S.,  1972, \mn@doi [\apjl]
  {10.1086/180994}, \href
  {https://ui.adsabs.harvard.edu/abs/1972ApJ...175L..99M} {175, L99}

\bibitem[\protect\citeauthoryear{{Nahar}}{{Nahar}}{1995}]{1995ApJS..101..423N}
{Nahar} S.~N.,  1995, \mn@doi [\apjs] {10.1086/192248}, \href
  {http://adsabs.harvard.edu/abs/1995ApJS..101..423N} {101, 423}

\bibitem[\protect\citeauthoryear{{Nahar}}{{Nahar}}{2000}]{2000ApJS..126..537N}
{Nahar} S.~N.,  2000, \mn@doi [\apjs] {10.1086/313307}, \href
  {http://adsabs.harvard.edu/abs/2000ApJS..126..537N} {126, 537}

\bibitem[\protect\citeauthoryear{{Nayfonov}, {D{\"a}ppen}, {Hummer}  \&
  {Mihalas}}{{Nayfonov} et~al.}{1999}]{1999ApJ...526..451N}
{Nayfonov} A.,  {D{\"a}ppen} W.,  {Hummer} D.~G.,   {Mihalas} D.,  1999,
  \mn@doi [\apj] {10.1086/307972}, \href
  {https://ui.adsabs.harvard.edu/abs/1999ApJ...526..451N} {526, 451}

\bibitem[\protect\citeauthoryear{{Nieva} \& {Przybilla}}{{Nieva} \&
  {Przybilla}}{2012}]{2012A&A...539A.143N}
{Nieva} M.-F.,  {Przybilla} N.,  2012, \mn@doi [\aap]
  {10.1051/0004-6361/201118158}, \href
  {http://adsabs.harvard.edu/abs/2012A%26A...539A.143N} {539, A143}

\bibitem[\protect\citeauthoryear{{O'brian} \& {Lawler}}{{O'brian} \&
  {Lawler}}{1991}]{1991PhRvA..44.7134O}
{O'brian} T.~R.,  {Lawler} J.~E.,  1991, \mn@doi [\pra]
  {10.1103/PhysRevA.44.7134}, \href
  {http://ads.ari.uni-heidelberg.de/abs/1991PhRvA..44.7134O} {44, 7134}

\bibitem[\protect\citeauthoryear{{Przybilla}, {Nieva}  \& {Butler}}{{Przybilla}
  et~al.}{2008}]{2008ApJ...688L.103P}
{Przybilla} N.,  {Nieva} M.-F.,   {Butler} K.,  2008, \mn@doi [\apjl]
  {10.1086/595618}, \href
  {https://ui.adsabs.harvard.edu/abs/2008ApJ...688L.103P} {688, L103}

\bibitem[\protect\citeauthoryear{{Przybilla}, {Nieva}  \& {Butler}}{{Przybilla}
  et~al.}{2011}]{2011JPhCS.328a2015P}
{Przybilla} N.,  {Nieva} M.-F.,   {Butler} K.,  2011, \mn@doi [Journal of
  Physics Conference Series] {10.1088/1742-6596/328/1/012015}, \href
  {http://ads.ari.uni-heidelberg.de/abs/2011JPhCS.328a2015P} {328, 012015}

\bibitem[\protect\citeauthoryear{{Ryabchikova}, {Piskunov}, {Kurucz},
  {Stempels}, {Heiter}, {Pakhomov}  \& {Barklem}}{{Ryabchikova}
  et~al.}{2015}]{2015PhyS...90e4005R}
{Ryabchikova} T.,  {Piskunov} N.,  {Kurucz} R.~L.,  {Stempels} H.~C.,  {Heiter}
  U.,  {Pakhomov} Y.,   {Barklem} P.~S.,  2015, \mn@doi [\physscr]
  {10.1088/0031-8949/90/5/054005}, \href
  {http://adsabs.harvard.edu/abs/2015PhyS...90e4005R} {90, 054005}

\bibitem[\protect\citeauthoryear{{Sadakane} \& {Nishimura}}{{Sadakane} \&
  {Nishimura}}{2017}]{2017PASJ...69...48S}
{Sadakane} K.,  {Nishimura} M.,  2017, \mn@doi [\pasj] {10.1093/pasj/psx024},
  \href {https://ui.adsabs.harvard.edu/abs/2017PASJ...69...48S} {69, 48}

\bibitem[\protect\citeauthoryear{{Sadakane} \& {Nishimura}}{{Sadakane} \&
  {Nishimura}}{2019}]{2019PASJ...71...45S}
{Sadakane} K.,  {Nishimura} M.,  2019, \mn@doi [\pasj] {10.1093/pasj/psz016},
  \href {https://ui.adsabs.harvard.edu/abs/2019PASJ...71...45S} {71, 45}

\bibitem[\protect\citeauthoryear{{Seaton}}{{Seaton}}{1962}]{1962amp..conf..375S}
{Seaton} M.~J.,  1962, in {Bates} D.~R.,  ed., Atomic and Molecular Processes.
  p.~375

\bibitem[\protect\citeauthoryear{{Seaton}}{{Seaton}}{1987}]{1987JPhB...20.6363S}
{Seaton} M.~J.,  1987, \mn@doi [Journal of Physics B Atomic Molecular Physics]
  {10.1088/0022-3700/20/23/026}, \href
  {https://ui.adsabs.harvard.edu/abs/1987JPhB...20.6363S} {20, 6363}

\bibitem[\protect\citeauthoryear{{Shchukina}, {Sukhorukov}  \& {Trujillo
  Bueno}}{{Shchukina} et~al.}{2012}]{2012ApJ...755..176S}
{Shchukina} N.,  {Sukhorukov} A.,   {Trujillo Bueno} J.,  2012, \mn@doi [\apj]
  {10.1088/0004-637X/755/2/176}, \href
  {https://ui.adsabs.harvard.edu/abs/2012ApJ...755..176S} {755, 176}

\bibitem[\protect\citeauthoryear{{Shi}, {Gehren}, {Butler}, {Mashonkina}  \&
  {Zhao}}{{Shi} et~al.}{2008}]{Shi_si_sun}
{Shi} J.~R.,  {Gehren} T.,  {Butler} K.,  {Mashonkina} L.~I.,   {Zhao} G.,
  2008, \mn@doi [\aap] {10.1051/0004-6361:200809452}, \href
  {http://adsabs.harvard.edu/abs/2008A%26A...486..303S} {486, 303}

\bibitem[\protect\citeauthoryear{{Shulyak}, {Tsymbal}, {Ryabchikova},
  {St{\"u}tz}  \& {Weiss}}{{Shulyak} et~al.}{2004}]{2004AA...428..993S}
{Shulyak} D.,  {Tsymbal} V.,  {Ryabchikova} T.,  {St{\"u}tz} C.,   {Weiss}
  W.~W.,  2004, \mn@doi [\aap] {10.1051/0004-6361:20034169}, 428, 993

\bibitem[\protect\citeauthoryear{{Singh}, {Aggarwal}, {Jha}, {Singh}  \&
  {Mohan}}{{Singh} et~al.}{2011}]{2011CaJPh..89.1119S}
{Singh} J.,  {Aggarwal} S.,  {Jha} A.~K.~S.,  {Singh} A.~K.,   {Mohan} M.,
  2011, \mn@doi [Canadian Journal of Physics] {10.1139/p11-106}, \href
  {http://adsabs.harvard.edu/abs/2011CaJPh..89.1119S} {89, 1119}

\bibitem[\protect\citeauthoryear{{Sitnova}, {Mashonkina}  \&
  {Ryabchikova}}{{Sitnova} et~al.}{2013}]{sitnova_o}
{Sitnova} T.~M.,  {Mashonkina} L.~I.,   {Ryabchikova} T.~A.,  2013, \mn@doi
  [Astronomy Letters] {10.1134/S1063773713020084}, \href
  {http://adsabs.harvard.edu/abs/2013AstL...39..126S} {39, 126}

\bibitem[\protect\citeauthoryear{{Sitnova}, {Mashonkina}  \&
  {Ryabchikova}}{{Sitnova} et~al.}{2016}]{sitnova_ti}
{Sitnova} T.~M.,  {Mashonkina} L.~I.,   {Ryabchikova} T.~A.,  2016, \mn@doi
  [\mnras] {10.1093/mnras/stw1202}, \href
  {http://adsabs.harvard.edu/abs/2016MNRAS.461.1000S} {461, 1000}

\bibitem[\protect\citeauthoryear{{Sitnova}, {Mashonkina}  \&
  {Ryabchikova}}{{Sitnova} et~al.}{2018}]{2018MNRAS.477.3343S}
{Sitnova} T.~M.,  {Mashonkina} L.~I.,   {Ryabchikova} T.~A.,  2018, \mn@doi
  [\mnras] {10.1093/mnras/sty810}, \href
  {https://ui.adsabs.harvard.edu/abs/2018MNRAS.477.3343S} {477, 3343}

\bibitem[\protect\citeauthoryear{{Smith}, {Huber}, {Tozzi}, {Griesinger},
  {Cardon}  \& {Lombardi}}{{Smith} et~al.}{1987}]{1987ApJ...322..573S}
{Smith} P.~L.,  {Huber} M.~C.~E.,  {Tozzi} G.~P.,  {Griesinger} H.~E.,
  {Cardon} B.~L.,   {Lombardi} G.~G.,  1987, \mn@doi [\apj] {10.1086/165752},
  \href {https://ui.adsabs.harvard.edu/abs/1987ApJ...322..573S} {322, 573}

\bibitem[\protect\citeauthoryear{{Tsymbal}, {Ryabchikova}  \&
  {Sitnova}}{{Tsymbal} et~al.}{2019}]{2019ASPC}
{Tsymbal} V.,  {Ryabchikova} T.,   {Sitnova} T.,  2019, in {Kudryavtsev} D.~O.,
   {Romanyuk} I.~I.,   {Yakunin} I.~A.,  eds,  Astronomical Society of the
  Pacific Conference Series Vol. 518, Astronomical Society of the Pacific
  Conference Series. pp 247--252

\bibitem[\protect\citeauthoryear{{Vernazza}, {Avrett}  \& {Loeser}}{{Vernazza}
  et~al.}{1976}]{1976ApJS...30....1V}
{Vernazza} J.~E.,  {Avrett} E.~H.,   {Loeser} R.,  1976, \mn@doi [\apjs]
  {10.1086/190356}, \href
  {https://ui.adsabs.harvard.edu/abs/1976ApJS...30....1V} {30, 1}

\bibitem[\protect\citeauthoryear{{Wedemeyer}}{{Wedemeyer}}{2001}]{2001A&A...373..998W}
{Wedemeyer} S.,  2001, \mn@doi [\aap] {10.1051/0004-6361:20010663}, \href
  {https://ui.adsabs.harvard.edu/abs/2001A&A...373..998W} {373, 998}

\bibitem[\protect\citeauthoryear{{Wiese}, {Smith}  \& {Miles}}{{Wiese}
  et~al.}{1969}]{1969atp..book.....W}
{Wiese} W.~L.,  {Smith} M.~W.,   {Miles} B.~M.,  1969, {Atomic transition
  probabilities. Vol. 2: Sodium through Calcium. A critical data compilation}

\bibitem[\protect\citeauthoryear{{van Regemorter}}{{van
  Regemorter}}{1962}]{Reg1962}
{van Regemorter} H.,  1962, \mn@doi [\apj] {10.1086/147445}, \href
  {http://ads.ari.uni-heidelberg.de/abs/1962ApJ...136..906V} {136, 906}

\makeatother
\end{thebibliography}
\bibliographystyle{mnras}

\label{lastpage}
\end{document}